\documentclass[conference, 10pt]{IEEEtran}
\usepackage{amsmath,amsthm,amssymb,amsfonts}
\usepackage[linesnumbered,ruled,vlined]{algorithm2e}
\usepackage{setspace}
\usepackage{bm}
\usepackage{threeparttable}
\usepackage{tabularx}
\usepackage{graphicx}
\usepackage{textcomp}
\usepackage{hhline}
\usepackage[hidelinks]{hyperref}
\usepackage[usenames,dvipsnames,table,xcdraw]{xcolor}
\usepackage{makecell}
\usepackage{subcaption}
\usepackage{enumitem}
\usepackage{float}
\usepackage{tikz}
\usepackage{pgfplots}
\usepackage{comment}
\usetikzlibrary{arrows,decorations,backgrounds,shapes,patterns,shadows,positioning,decorations.pathmorphing,matrix,calc,automata}

\newtheorem{theorem}{Theorem}[section]
 
\newtheorem{lemma}[theorem]{Lemma}
\newtheorem{observation}{Observation}[section]
\theoremstyle{definition}
\newtheorem{definition}{Definition}[section]
\newtheorem{example}{Example}[section]

\newif\iffinal
\finaltrue

\usepackage{todonotes}

\iffinal
    \newcommand{\remove}[1]{}
    \newcommand{\removeeq}[2]{}
    \newcommand{\revise}[1]{#1}
\else
    \newcommand{\remove}[1]{{\color{red}\sout{#1}}}
    \newcommand{\removeeq}[2]{{\color{red}\begin{equation}\text{\sout{\ensuremath{#2}\tag{\sout{#1}}}}\end{equation}}}
    \newcommand{\revise}[1]{{\color{blue}#1}}
\fi

\definecolorseries{test}{rgb}{grad}[rgb]{.95,.55,.55}{11,11,17}
\resetcolorseries[10]{test}
\newcommand{\addtodoeditor}[1]{%
    \colorlet{#1}{test!!+!50}
    \expandafter\newcommand\csname#1\endcsname [1]{%
        \todo[color=#1,size=\tiny]{\sffamily\textbf{\uppercase{#1}:}
    ##1}\xspace%
    }
    \expandafter\newcommand\csname#1i\endcsname [1]{%
        \todo[inline, color=#1]{\sffamily\textbf{\uppercase{#1}:} ##1}\xspace%
    }
}

\addtodoeditor{bs}
\addtodoeditor{tk}

\newcommand{\eg}{{\it e.g.}\xspace}
\newcommand{\ie}{{\it i.e.}\xspace}

\let\oldnl\nl
\newcommand{\nonl}{\renewcommand{\nl}{\let\nl\oldnl}}

\def\BibTeX{{\rm B\kern-.05em{\sc i\kern-.025em b}\kern-.08em
    T\kern-.1667em\lower.7ex\hbox{E}\kern-.125emX}}

\begin{document}

\title{Strict Partitioning for Sporadic Rigid Gang Tasks}

\author{
\IEEEauthorblockN{
Binqi Sun\IEEEauthorrefmark{1},
Tomasz Kloda\IEEEauthorrefmark{2},
Marco Caccamo\IEEEauthorrefmark{1}
}
\IEEEauthorblockA{\IEEEauthorrefmark{1}Technical University of Munich, Germany}
\IEEEauthorblockA{\IEEEauthorrefmark{2}LAAS-CNRS, Université de Toulouse, INSA, Toulouse, France
\\Email: binqi.sun@tum.de, tkloda@laas.fr, mcaccamo@tum.de}
}

\IEEEoverridecommandlockouts
\IEEEpubid{\makebox[\columnwidth]{© 2024 IEEE. Citation information: DOI 10.1109/RTAS61025.2024.00028 \hfill}
\hspace{\columnsep}\makebox[\columnwidth]{ }}

\maketitle

\IEEEpubidadjcol

\begin{abstract}  
The \textit{rigid gang} task model is based on the idea of executing multiple threads simultaneously on a \textit{fixed} number of processors to increase efficiency and performance.
Although there is extensive literature on global rigid gang scheduling, partitioned approaches have several practical advantages (\eg, task isolation and reduced scheduling overheads).
In this paper, we propose a new partitioned scheduling strategy for rigid gang tasks, named \textit{strict partitioning}.
The method creates disjoint partitions of tasks and processors to avoid inter-partition interference. 
Moreover, it tries to assign tasks with similar volumes (\ie, parallelisms) to the same partition so that the intra-partition interference can be reduced. 
Within each partition, the tasks can be scheduled using any type of scheduler, which allows the use of a less pessimistic schedulability test. 
Extensive synthetic experiments and a case study based on Edge TPU benchmarks show that strict partitioning achieves better schedulability performance than state-of-the-art global gang schedulability analyses for both preemptive and non-preemptive rigid gang task sets. 
\end{abstract}


\section{Introduction}
\label{sec:intro}

Gangs are fine-grained parallel jobs that have to start at the same time and execute in parallel across multiple processing units. 
Gang scheduling has been widely used in high-performance computing~\cite{Ousterhout:1982,Zhang:2000}, distributed cloud~\cite{Moschakis:2010}, and containers~\cite{Carrion:2022}.
In recent years, it is also increasingly gaining popularity in embedded systems. 
Especially, gang scheduling is well-suited for emerging deep-learning applications~\cite{Dong:2021,Bian:2022} deployed on highly parallel hardware accelerators~\cite{Seshadri:2022,Micaela:2020} that require significant computational resources and may involve complex inter-task communication and dependencies.

Most of the approaches for gang scheduling in real-time systems are \emph{global}~\cite{Dong:2021,Dong:2022,Lee:2022,Kim:2016,Binqi:2023,lee2022response,Nelissen:2022,Kato:2009,Goossens:2010,Goossens:2016,Dong:2017}, wherein
tasks are not pinned to any particular processor and can start execution on any available processors.
Despite the relatively lesser interest in \emph{partitioned} scheduling~\cite{Ueter:2021} (\emph{i.e.,}~tasks are statically allocated to individual processors), such type of scheduling can offer benefits in specific scenarios where task isolation and reduced scheduling overheads~\cite{Andrea:2010,Lelli:2012,Brandenburg:2016} are \mbox{essential~requirements.}

\begin{figure}[t]
\centering
\includegraphics[width=0.85\linewidth]{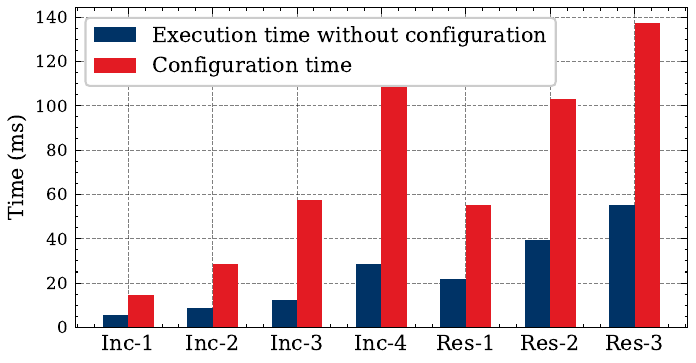}
\caption{Configuration time vs net execution time of DNN inferences on 16 Edge TPUs. ``Inc-1,2,3,4" denote Inception-v1~\cite{inception_v1_2014}, v2~\cite{inception_v2_v3_2016}, v3~\cite{inception_v2_v3_2016}, v4~\cite{inception_v4_inception_resnet_V2_2018}, respectively; ``Res-1,2,3" denote ResNet-50,101,152~\cite{resnetv1_50_101_152_2015}, respectively.}
\label{fig:config_time}
\end{figure}

The migration cost is one of the major drawbacks of global scheduling. 
Parallel tasks are often deployed on hardware accelerators where the migration and setup times are usually high (\emph{e.g.,} \emph{FPGA} dynamic partial reconfiguration~\cite{Biondi:2017} or \emph{Edge TPU} model parameter loading~\cite{Binqi:2023,Han:2023}).
Figure~\ref{fig:config_time} shows the configuration and the net execution times of seven representative deep neural networks (DNNs) on an AI Accelerator card integrated with 16~Edge TPUs.
Each Edge TPU has 8 MB on-chip memory and can be used to store DNN model weights. 
Partitioned scheduling reserves processing units for each model so that the model weights can be cached in the internal memory statically, eliminating the need for frequent and costly memory allocation during runtime.

The current state-of-the-art multiprocessor analyses for sequential tasks under global scheduling rely primarily on sufficient-only conditions~\cite{Sun:2018}.
Their pessimism~\cite{Baker:2006} is due to the fact that synchronous arrival sequence is not necessarily the worst-case scenario~\cite{Lauzac:1998, Baruah:2010}. Global gang scheduling, besides the same old problems, is burdened with interference overestimation resulting from the difference in gang tasks' parallelism levels (\emph{e.g.,}~finding the set of interfering jobs that may exactly fit on the available processors~\cite{Lee:2022} and unbounded priority-inversion in non-preemptive gang scheduling~\cite{Dong:2019}).

While partitioned scheduling can help reduce runtime overheads,
the task allocation is equivalent to the bin-packing problem~\cite{Sarkar:1987} and is hence highly intractable (NP-hard in the strong sense~\cite{Johnson:1974}).
As different gang tasks may have different parallelism levels, the gang partitioning is closely related to the two-dimensional bin packing problem (\emph{i.e.,}~packing a set of rectangular items into rectangular bins)~\cite{Baker:1980}. Thus, the tractability challenge extends to the gang partitioning problem.
Moreover, partitioned systems of sequential tasks can be easily analyzed using exact uniprocessor schedulability tests. However, for parallel tasks, more complicated tests may be necessary to account for the anomalies that may occur in partitioned gang systems where different gang task groups share some, but not all, processors (see Section~\ref{sec:motivate})~\cite{Ueter:2021}.

This paper proposes a simple yet effective method, named \emph{strict partitioning}, for scheduling rigid gang tasks (\emph{i.e.,} gang tasks with fixed parallelism levels) on identical multiprocessor platforms. It builds disjoint partitions of tasks and processors so that the tasks in different partitions do not interfere with each other.
Within the boundaries of each partition, tasks can run under any type of online scheduler. Moreover, strict partitioning tries to group tasks with similar parallelism levels onto the same partition so that uniprocessor scheduler (\emph{e.g.,~Deadline Monotonic (DM)~\cite{Audsley:1991}, Earliest Deadline First (EDF)}~\cite{Liu:1973}) and exact schedulability tests are applicable.

The contributions are summarized as follows.
\begin{itemize}
    \item We propose a new \textit{strict partitioning} strategy for scheduling rigid gang tasks.
    \item We propose a \textit{first-fit decreasing volume (FFDV)} heuristic to partition rigid gang tasks and multiprocessor platforms.
    \item We present two strict partitioning variants \texttt{SP-U} and \texttt{SP-G} by combining FFDV with different online schedulers. For \texttt{SP-U}, we prove utilization bounds. For \texttt{SP-G}, we improve FFDV to achieve better performance.
    \item We evaluate the proposed strategy and algorithms by comparing them with state-of-the-art preemptive and non-preemptive gang scheduling techniques on synthetic task sets and a case study based on Edge TPU benchmarks. 
\end{itemize}

The remainder of the paper is organized as follows. Relevant previous works are presented in Section~\ref{sec:related_work}. Section~\ref{sec:strict_partition} describes the proposed strict partitioning strategy, and Section~\ref{sec:motivate} compares strict partitioning with existing gang scheduling methods through illustrative examples. We propose strict partitioning algorithms in Section~\ref{sec:alg} and present evaluation results in Section~\ref{sec:exp}. Section~\ref{sec:conclusion} concludes the paper.

\section{Related Work}
\label{sec:related_work}

Three main paradigms are adopted to schedule real-time tasks on a multiprocessor~\cite{Baruah:2015_book,Davis:2011}: \emph{global} (each job can execute upon any processor), \emph{partitioned} (each job is mapped to an individual processor), and \emph{semi-partitioned} (several tasks are split into subtasks assigned to different processors).

\textit{Partitioned scheduling for sequential tasks.}
The partitioned schedulers incur low runtime overheads  but require solving a
task-to-processor bin-packing problem.
Since the partitioning problem is NP-hard in the strong sense~\cite{ekberg2021partitioned}, many approximation methods (based on bin-packing heuristics) for sequential preemptive tasks were proposed~\cite{Burchard:1995,Dhall:1978,Lopez:2000,Lopez:2004, Baruah:2005,baruah2006partitioned,Fisher:2006_2}.
Other works formulate the partitioning as an integer
linear programming (ILP) problem~\cite{Baruah:2008,Zheng:2007}.
The partitioning methods for non-preemptive scheduling~\cite{Fisher:2006,Berna:2012,Mayank:2017} try to cluster on the same processor the tasks with similar period ranges to reduce the blocking factor.
\emph{Semi-partitioned} approaches reclaim spare processing capacity in partitioned
systems by splitting certain tasks among processors~\cite{Burns:2010,Andersson:2006}.

\textit{Partitioned scheduling for parallel tasks.}
Casini et al.~\cite{Casini:2018} and Wu et al.~\cite{Wu:2023} propose partitioning algorithms for parallel tasks modeled as a \emph{directed acyclic graph (DAG)}.
Zahaf et al.~\cite{Zahaf:2010} consider the allocation of conditional DAGs to remove unnecessary preemptions.
\textit{Federated scheduling} categorizes parallel tasks, according to their utilization factor, into heavy tasks and light tasks~\cite{li2014analysis,Baruah:2015}. Each heavy task is assigned a number of dedicated processors. The remaining processors are assigned to all light tasks, which are executed as sequential tasks. In~\cite{Jiang:2021,Ueter:2018}, heavy tasks are allowed to share processors, and in~\cite{jiang2017semi}, they can partially execute on the processors from the light tasks' pool.
However, unlike DAG tasks, whose subtasks can run on a different number of processors and do not have to start execution synchronously, the parallelism level of a rigid gang task cannot be changed at runtime. Thus, these partitioning methods designed for DAG tasks cannot be applied to rigid gang scheduling. The major concern of partitioning rigid gang tasks is how to effectively group tasks with fixed parallelism levels that are not necessarily the same.

\textit{Gang scheduling.}
The problem of rigid gang scheduling was shown to be NP-hard~\cite{kubale1987complexity}.
Several works have exploited the idea of \emph{proportionate progress}~\cite{Baruah:1995,Baruah:1993} to derive the optimal scheduling algorithms for periodic implicit deadline \emph{rigid}~\cite{Goossens:2016} and \emph{malleable}~\cite{Collette}  (\emph{i.e.,} job parallelism level can be adjusted at runtime) gang tasks.
Other streams of works studied rigid gang tasks under \emph{global fixed-priority} preemptive~\cite{Berten:2011,lee2022response} and non-preemptive~\cite{Binqi:2023,Kim:2016,Lee:2022}, as well as  \emph{global EDF} preemptive~\cite{Kato:2009,Dong:2021,Dong:2017,Richard:2017} and non-preemptive  ~\cite{Dong:2019,Dong:2022} scheduling.
Some recent works have also considered moldable gang tasks~\cite{Nelissen:2022}.
In this work, we apply a partitioned approach and compare its schedulability performance with the above state-of-the-art global policies in Section~\ref{sec:exp}.

\emph{Gang partitioning.}
Ueter et al.~\cite{Ueter:2021} proposed the concept of stationary scheduling and a task-to-processor allocation heuristic for rigid gangs. In stationary scheduling, each gang task is mapped to a set of processors such that the number of processors equals the task parallelism level. The schedulability problem is equivalent to the uniprocessor schedulability of self-suspending  tasks~\cite{Chen:2019}.
If tasks $\tau_i$ and $\tau_j$ run on the same processors but $\tau_j$ runs also on other processors where another task $\tau_k$ is allocated to, then $\tau_k$ can interfere indirectly with the $\tau_i$'s execution ($\tau_k \rightarrow \tau_j \rightarrow \tau_i $).
In this work, we propose a different partitioning approach where tasks cannot run on the processors allocated to the tasks from different partitions. 
This eliminates the interference between the partitions and simplifies the scheduling problem, enabling the use of exact schedulability tests. On the other hand, more constrained processor allocation can lead to less efficient utilization of processing capacity.  
In Section~\ref{sec:exp}, we compare both approaches and identify their respective advantages.

\section{Strict Partitioning for Rigid Gang Tasks} 
\label{sec:strict_partition}

\subsection{Task and Platform Model}
We consider a multiprocessor platform with a set $\Pi$ of $M$ identical processors. A task set $\tau$ is composed of $n$ independent sporadic \textit{rigid gang} tasks executing on $\Pi$. Each task~$\tau_i \in \tau$ ($1\leq i\leq n$) gives rise to an infinite sequence of jobs with consecutive jobs' invocations (arrivals) separated by at least~$T_i$ time units (\emph{i.e.,} \emph{sporadic} task). 
We use $J_i$ to denote a job of task~$\tau_i$. Job~$J_i$ released at time (arrival time) $r_i$ has an absolute deadline~\mbox{$r_i+D_i$} and must complete its execution by that time where~$D_i\leq T_i$ (\emph{i.e.,} \emph{constrained} deadlines). 
Each job of task~$\tau_i$ executes simultaneously on~$m_i$ processors for at most~$C_i$ time units ($m_i$ is called \textit{task volume} or \textit{parallelism} interchangeably). As a result, a gang task can be characterized by a 4-tuple~$\tau_i = (C_{i}, T_i, D_i,m_i)$.
We assume that all the above parameters are non-negative integers. The tasks do not self-suspend and cannot be blocked by other tasks other than due to contention on processors.
Additionally,
we define the \emph{sequential utilization} of task~$\tau_i$ as  $u_i = C_i / T_i$ and its utilization as $U_{i} = {m_i \cdot C_{i}}  / T_i = m_i \cdot u_i$.
The task set utilization is the sum of task utilizations,~$U = \sum_{i=1}^n{U_{i}}$.
Moreover, 
$\overline{m}$ and $\underline{m}$ denote the maximum and minimum task volume among all the tasks in the task set.

\subsection{Strict Partitioning Strategy}
The strict partitioning strategy includes two aspects: \textit{\mbox{offline partitioning}} and \textit{online scheduling}.

\subsubsection{Offline Partitioning}

Offline partitioning involves dividing processors into disjoint partitions and statically allocating tasks to these partitions.

\begin{definition}[Strict Partitioning]
Given a set of identical processors $\Pi$ and a set of rigid gang tasks $\tau$, its \emph{strict partitioning} is such assignment where processors $\Pi$ are divided into disjoint subsets $\rho = \{\rho_j \subseteq \Pi, \forall j\}$ and each task $\tau_k \in \tau$ is assigned to one and only one partition.
\end{definition}
The set of tasks assigned to processor partition $\rho_j$ is denoted by $\tau(\rho_j) \subseteq \tau$ and
the volume of partition $\rho_j$ (\ie, the number of processors allocated to $\rho_j$)  by $|\rho_j|$.

\subsubsection{Online Scheduling}
Once the offline partitioning is determined, the tasks allocated to each processor partition will be scheduled by an \textit{online} real-time scheduler at runtime. Both \textit{uniprocessor task schedulers} (\eg,~uniprocessor \emph{DM}~\cite{Audsley:1991} and \emph{EDF}~\cite{Liu:1973}) and \textit{global gang task schedulers} (\eg,~global \emph{FP}~\cite{Lee:2022,Binqi:2023} and \emph{EDF}~\cite{lee2022response} gang schedulers) can be applied. When a uniprocessor task scheduler is used, only one job is allowed to execute on the processor partition at each time instant (\ie,~different jobs cannot run in parallel). By contrast, a global gang task scheduler allows tasks to start their execution whenever a sufficient number of processors are available within the partition.

\section{Illustrative Examples: Comparative Study} 
\label{sec:motivate}

 We will illustrate the tradeoffs that arise when considering different approaches to rigid gang scheduling.
 It includes the pessimism of global fixed-priority scheduling, a comparison between  \emph{stationary} and \emph{strict} partitioning of rigid gangs upon multiprocessor, and priority-inversion in non-preemptive global~policies. In what follows, we assume that the tasks are 
scheduled by a fixed-priority scheduler and 
indexed in decreasing priority order (\emph{i.e.,}~task~$\tau_1$ has the highest priority).

\subsection{Pessimisms in Global Gang Schedulability Analyses}
\label{sec:eg_global}
We recall two phenomena in global gang schedulability analyses identified by~Lee et al.~\cite{lee2022response}.
\subsubsection{Non-parallel Execution}
Classic multiprocessor frameworks for response time analysis of sequential tasks~\cite{Marko:2007,Marko:2009} consider that every higher-priority task can interfere with the execution of the task under analysis, regardless of the execution of other interfering tasks.
In rigid gang scheduling, certain tasks cannot run at the same time if there are not enough processors for all of them~\cite{lee2022response}.

\begin{example}
Let us consider three sporadic rigid gang tasks 
$\tau_1=(4,10,10,4)$, $\tau_2=(3,10,10,3)$, and $\tau_3=(5,10,10,1)$ running on five, $M{=}5$, identical processors~$P_{1-5}$. We check task~$\tau_3$'s schedulability. 
Although tasks $\tau_1$ and $\tau_2$ have higher priorities, they cannot interfere with $\tau_3$'s execution. 
Given their parallelism levels, $\tau_1$ and $\tau_2$ cannot execute at the same time, $m_1+m_2=4+3=7>M=5$, and there is always at least one processor for~$\tau_3$.
Figure~\ref{fig:global_11} shows a sample schedule.
\begin{figure}[hbt]

\begin{subfigure}[b]{0.24\textwidth}
    \begin{tikzpicture}[xscale=0.5,yscale=0.5]
    
    \draw[->] (-0.1,-0.15) -- (7.25,-0.15);
    \draw[->] (-0.2,-0.15) -- (-0.2,5.3);
    
    \foreach \x in {0,...,7}
        \draw (\x,-0.20) -- (\x,-0.1) node[yshift=-0.25cm,scale=0.6] {$\x$};
    
    \draw (-0.3,3.5) -- (-0.1,3.5) node[xshift=-0.2cm,scale=0.6] {$P_2$};
    \draw (-0.3,4.5) -- (-0.1,4.5) node[xshift=-0.2cm,scale=0.6] {$P_1$};
    \draw (-0.3,0.5) -- (-0.1,0.5) node[xshift=-0.2cm,scale=0.6] {$P_5$};
    \draw (-0.3,1.5) -- (-0.1,1.5) node[xshift=-0.2cm,scale=0.6] {$P_4$};
    \draw (-0.3,2.5) -- (-0.1,2.5) node[xshift=-0.2cm,scale=0.6] {$P_3$};

	\draw[pattern color=red!60,pattern=north west lines] (4,2.0) rectangle (7.0,5.0) node[pos=.5,scale=0.75] {$\tau_2$};

	\draw[pattern color=black!60,pattern=north west lines] (0,0.0) rectangle (5.0,1.0) node[pos=.5,scale=0.75] {$\tau_3$};

    \draw[pattern color=blue!60,pattern=north west lines] (0,1.0) rectangle (4.0,5.0) node[pos=.5,scale=0.75] {$\tau_1$};
    \end{tikzpicture}
        \caption{Non-parallel execution}
    \label{fig:global_11}
\end{subfigure}
\begin{subfigure}[b]{0.24\textwidth}
    \begin{tikzpicture}[xscale=0.5,yscale=0.5]
    
    \draw[->] (-0.1,-0.15) -- (7.25,-0.15);
    \draw[->] (-0.2,-0.15) -- (-0.2,5.3);
    
    \foreach \x in {0,...,7}
        \draw (\x,-0.20) -- (\x,-0.1) node[yshift=-0.25cm,scale=0.6] {$\x$};
    
    \draw (-0.3,3.5) -- (-0.1,3.5) node[xshift=-0.2cm,scale=0.6] {$P_2$};
    \draw (-0.3,4.5) -- (-0.1,4.5) node[xshift=-0.2cm,scale=0.6] {$P_1$};
    \draw (-0.3,0.5) -- (-0.1,0.5) node[xshift=-0.2cm,scale=0.6] {$P_5$};
    \draw (-0.3,1.5) -- (-0.1,1.5) node[xshift=-0.2cm,scale=0.6] {$P_4$};
    \draw (-0.3,2.5) -- (-0.1,2.5) node[xshift=-0.2cm,scale=0.6] {$P_3$};

    \draw[pattern color=blue!60,pattern=north west lines] (0,2.0) rectangle (7.0,5.0) node[pos=.5,scale=0.75] {$\tau_1$};
    
	\draw[pattern color=red!60,pattern=north west lines] (0,1.0) rectangle (3.0,2.0) node[pos=.5,scale=0.75] {$\tau_2$};

	\draw[pattern color=violet!60,pattern=north west lines] (3.0,0.0) rectangle (6.0,2.0) node[pos=.5,scale=0.75] {$\tau_3$};

    \draw[pattern color=black!60,pattern=north west lines] (6.0,0.0) rectangle (7.0,2.0) node[pos=.5,scale=0.75] {$\tau_4$};
    \end{tikzpicture}
        \caption{Overestimation of interference}
    \label{fig:global_12}
\end{subfigure}
    \caption{New sources of pessimism in global gang scheduling.}
    \vspace{-0.5em}
    \label{fig:global_2}
    \end{figure}
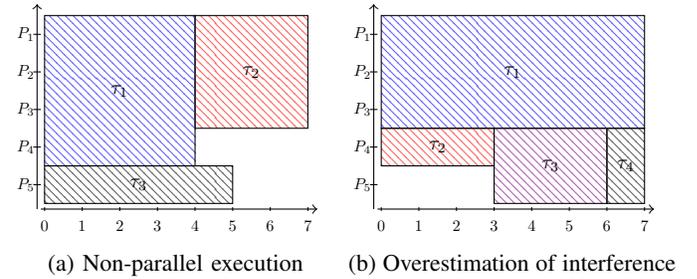
\end{example}

In general, identifying all sets of non-parallel executing jobs can be solved as the subset sum problem (\ie, find a subset of the integers that sum precisely to a given target value)~\cite{Kleinberg:2005}.
The problem is known to be NP-hard.
A heuristic is proposed in~\cite{lee2022response} to address such non-parallel execution constraints. 

\subsubsection{Interference Overestimation}
Classic multiprocessor frameworks for response time analysis of sequential tasks~\cite{Marko:2007,Marko:2009} consider the interference when all $M$ processors are occupied by the interfering tasks. In gang scheduling, task~$\tau_k$ cannot execute when at least $M-m_k+1$ processors are busy.
This complicates the interference estimation as a part of the interfering tasks can be executed on a different number of processors.
Such surplus interference can be reduced by deducting the overestimation by a heuristic proposed in~\cite{lee2022response}. However, computing the maximum overestimation deduction (\ie, finding the optimal $\tau'$ in Theorem 5 of \cite{lee2022response}) requires solving a knapsack problem, which is again NP-hard.

\begin{example}
Let us consider four sporadic rigid gang tasks
$\tau_1=(7,10,10,3)$, $\tau_2=(3,10,10,1)$, $\tau_3=(3,10,10,2)$, and $\tau_4=(1,10,10,2)$ running on five, $M{=}5$, identical processors~$P_{1-5}$. We check task $\tau_4$'s schedulability. To do that, we need to quantify the interference of higher-priority tasks. These tasks interfere with $\tau_4$ when they run on at least four~processors.
Figure~\ref{fig:global_12} illustrates the interference computation.
When  $\tau_1$ and $\tau_2$ run concurrently in interval $[0,3)$, four processors are busy, and the workload of interfering tasks is $4\cdot 3{=}12$.
When $\tau_1$ and $\tau_3$ run concurrently in interval $[3,6)$, five processors are busy, and the workload of interfering tasks is $5\cdot 3{=}15$.
Both intervals have equal lengths, and bigger interference during $[3,6)$ does not affect the task under study~$\tau_4$ more than smaller interference during interval~$[0,3)$.
\end{example}

While the heuristics proposed in~\cite{lee2022response} can mitigate the aforementioned pessimisms, the experimental results in Section~\ref{sec:exp} suggest that these pessimisms still exist in the analyses since it is not computationally tractable to solve them optimally. 

\subsection{Stationary Scheduling and Strict Partitioning}
\label{sec:eg_stationary}

In \emph{stationary scheduling}~\cite{Ueter:2021},  task $\tau_k$ with parallelism level~$m_k$ is allocated to $m_k$ processors. 
Some of these processors can be allocated to different tasks. If these tasks are also allocated to processors other than those shared with $\tau_k$, then they may transfer the interference from these processors to the processors shared with $\tau_k$.  
Such interfering behavior can be modeled and over-approximated by the
corresponding sequential tasks with dynamic \emph{self-suspension}~\cite{Chen:2019}.
In contrast, the strict partitioning proposed in this paper forbids overlapping in processor allocation and thus avoids scheduling anomalies related to the self-suspension task model.

\begin{example}
Consider a system comprising three sporadic rigid gang tasks 
$\tau_1=(2,5,5,1)$, $\tau_2=(3,6,6,2)$, and $\tau_3=(2,7,7,2)$ running on three identical processors $P_1$, $P_2$, and~$P_3$.
Figure~\ref{fig:stationary_vs_strict} shows task-to-processor assignments done by (a)~stationary gang heuristic and (b)~strict partitioning heuristic.
To avoid the combinatorial explosion, each heuristic uses a specific strategy. 
The stationary heuristic assigns the tasks to processors in a greedy way to have possibly the best processor utilization. The strict partitioning favors volume homogeneity within the same partition.
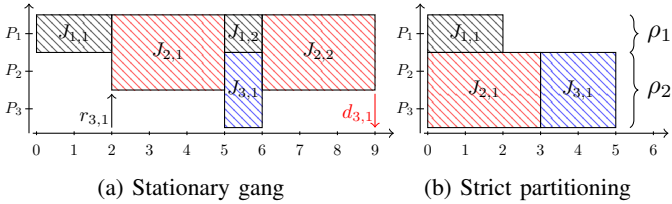
\begin{figure}[hbt]
\centering
\begin{subfigure}[b]{0.28\textwidth}
    \begin{tikzpicture}[xscale=0.5,yscale=0.5]
    
    \draw[->] (-0.1,-0.15) -- (9.5,-0.15);
    \foreach \x in {0,...,9}
        \draw (\x,-0.20) -- (\x,-0.1) node[yshift=-0.25cm,scale=0.6] {$\x$};

    \draw[->] (-0.2,-0.15) -- (-0.2,3.3);
    \draw (-0.3,0.5) -- (-0.1,0.5) node[xshift=-0.25cm,scale=0.6] {$P_3$};
    \draw (-0.3,1.5) -- (-0.1,1.5) node[xshift=-0.25cm,scale=0.6] {$P_2$};
    \draw (-0.3,2.5) -- (-0.1,2.5) node[xshift=-0.25cm,scale=0.6] {$P_1$};

	\draw[pattern color=black!60,pattern=north west lines] (0,3.0) rectangle (2.0,2.0) node[pos=.5,scale=0.75] {$J_{1,1}$};

	\draw[pattern color=black!60,pattern=north west lines] (5.00,3.0) rectangle (6.00,2.0) node[pos=.5,scale=0.75] {$J_{1,2}$};

    \draw[pattern color=red!60,pattern=north west lines] (2,1.0) rectangle (5.0,3.0) node[pos=.5,scale=0.75] {$J_{2,1}$};

	\draw[pattern color=red!60,pattern=north west lines] (6,1.0) rectangle (9.0,3.0) node[pos=.5,scale=0.75] {$J_{2,2}$};

	\draw[pattern color=blue!60,pattern=north west lines] (5,0.0) rectangle (6,2.0) node[pos=.5,scale=0.75] {$J_{3,1}$};

   \draw[->] (2.0,0.0) -- (2.0,0.90) node[pos=.25,scale=0.75,xshift=-0.3cm] {$r_{3,1}$};

   \draw[<-,red] (9,0.0) -- (9,0.90) node[pos=.4,scale=0.75,xshift=-0.3cm] {$d_{3,1}$};     
    
    \end{tikzpicture}
    \caption{Stationary gang}
    \label{fig:stationary_vs_strict_stationary}
    \end{subfigure}
    \begin{subfigure}[b]{0.20\textwidth}
    \begin{tikzpicture}[xscale=0.5,yscale=0.5]

    \draw[->] (-0.1,-0.15) -- (6.40,-0.15);
    \foreach \x in {0,...,6}
        \draw (\x,-0.20) -- (\x,-0.1) node[yshift=-0.25cm,scale=0.6] {$\x$};
        
    \draw[->] (-0.2,-0.15) -- (-0.2,3.3);
    \draw (-0.3,0.5) -- (-0.1,0.5) node[xshift=-0.25cm,scale=0.6] {$P_3$};
    \draw (-0.3,1.5) -- (-0.1,1.5) node[xshift=-0.25cm,scale=0.6] {$P_2$};
    \draw (-0.3,2.5) -- (-0.1,2.5) node[xshift=-0.25cm,scale=0.6] {$P_1$};
       
    \draw [decorate,decoration={brace,amplitude=3pt},xshift=5pt,yshift=0pt]  (5.2,3.0) -- (5.2,2.0) node [black,midway,xshift=0.4cm] {$\rho_1$}; 
    
    \draw [decorate,decoration={brace,amplitude=3pt},xshift=5pt,yshift=0pt]  (5.2,2.0) -- (5.2,0.0) node [black,midway,xshift=0.4cm] {$\rho_2$}; 
      
   	\draw[pattern color=black!60,pattern=north west lines] (0,2.0) rectangle (2.0,3.0) node[pos=.5,scale=0.75] {$J_{1,1}$};
   
    \draw[pattern color=red!60,pattern=north west lines] (0,0) rectangle (3.0,2.0) node[pos=.5,scale=0.75] {$J_{2,1}$};
      
    \draw[pattern color=blue!60,pattern=north west lines] (3,0) rectangle (5,2) node[pos=.5,scale=0.75] {$J_{3,1}$};
   
    \end{tikzpicture}
    \caption{Strict partitioning}
    \label{fig:stationary_vs_strict_strict}
    \end{subfigure}
    \caption{Strict partitioning better than stationary scheduling.}
    \label{fig:stationary_vs_strict}
    \end{figure} 

In the presented example, the stationary heuristic would allocate $\tau_1$ to $P_1$ and then $\tau_2$ to $P_1$ and $P_2$, as $\tau_1$ and $\tau_2$ can be both schedulable on $P_1$.
In contrast, the strict partitioning, to avoid mixing different parallelism levels, would open a new partition $\rho_2=\{P_2,P_3\}$ for $\tau_2$ ($\tau_2$ has a different volume as $\tau_1$ running on~$P_1$). 
This choice will be critical when assigning~$\tau_3$.

In the stationary gang assignment (a), since the utilization of $P_1$ is at $90\%$ and it cannot accommodate $\tau_3$ having sequential utilization of $u_3{=}28.5\%$, the heuristic allocates $\tau_3$ to $P_2$ (current utilization of $50\%$) and $P_3$ (idle). However, $\tau_3$ is not schedulable on $P_2$ and $P_3$ as $\tau_2$ exhibits self-suspending behavior~\cite{Chen:2019} due to $\tau_1$ preempting $\tau_2$ on~$P_1$. 
Task $\tau_3$'s first job, $J_{3,1}$, can miss its deadline when released at~$r_{3,1}{=}2$ while
tasks $\tau_1$ and $\tau_2$ release their first jobs, $J_{1,1}$ and $J_{2,1}$, at~$0$.
The $\tau_2$'s second instance, $J_{2,2}$, is released at $6$ and, if the $\tau_1$'s second instance, $J_{1,2}$, completes sooner than its worst-case execution time, $J_{2,2}$ will start its execution immediately on $P_1$ and $P_2$ occupying both processors until $9$.
Consequently, $J_{3,1}$ will not meet its deadline.

In the strict partitioning assignment (b), $\tau_3$ is allocated to the partition $\rho_2=\{P_1,P_2\}$ which, at the moment of assignment, is comprised of $\tau_2$ only (each processor has utilization of~$50\%$). 
Tasks $\tau_2$ and $\tau_3$ have both a volume of two and thus processors $P_2$ and $P_3$ are seen as a single resource. The tasks behave like on a traditional uniprocessor platform and are schedulable, showing no anomalies.
\end{example}

However, strict partitioning might lead to processor~under-utilization when tasks have very different parallelism levels.

\begin{example}
Consider a system comprising three sporadic rigid gang tasks 
$\tau_1=(1,3,3,1)$, $\tau_2=(1,4,4,2)$, and $\tau_3=(3,5,5,1)$ running on two identical processors $P_1$ and~$P_2$.
Figure~\ref{fig:strict_vs_stationary} shows task-to-processor assignments done by (a)~stationary gang heuristic and (b)~strict partitioning heuristic. 
In stationary gangs, sequential tasks are assigned to two different processors.
Despite the self-suspension phenomenon still present, all tasks are schedulable (task~$\tau_3$'s critical instant is at $r_{3,1}{=}1$ when other tasks start at $0$). 
Strict partitioning, because of the $\tau_2$'s volume, creates only one partition for all the tasks.
The problem simplifies to uniprocessor scheduling and the potential parallelism cannot be exploited.
Task $\tau_3$ released at $r_{3,1}{=}0$ misses its deadline at $d_{3,1}{=}5$.

\begin{figure}[hbt]
\centering
\begin{subfigure}[b]{0.20\textwidth}
    \begin{tikzpicture}[xscale=0.5,yscale=0.5]
    
    \draw[->] (-0.1,-0.15) -- (6.5,-0.15);
    \foreach \x in {0,...,6}
        \draw (\x,-0.20) -- (\x,-0.1) node[yshift=-0.25cm,scale=0.6] {$\x$};

    \draw[->] (-0.2,-0.15) -- (-0.2,2.3);
    \draw (-0.3,0.5) -- (-0.1,0.5) node[xshift=-0.25cm,scale=0.6] {$P_2$};
    \draw (-0.3,1.5) -- (-0.1,1.5) node[xshift=-0.25cm,scale=0.6] {$P_1$};

	\draw[pattern color=black!60,pattern=north west lines] (0,1.0) rectangle (1.0,2.0) node[pos=.5,scale=0.75] {$J_{1,1}$};

	\draw[pattern color=red!60,pattern=north west lines] (1.00,0.0) rectangle (2.00,2.0) node[pos=.5,scale=0.75] {$J_{2,1}$};

    \draw[pattern color=black!60,pattern=north west lines] (3,1.0) rectangle (4.0,2.0) node[pos=.5,scale=0.75] {$J_{1,2}$};

	\draw[pattern color=red!60,pattern=north west lines] (4,0.0) rectangle (5.0,2.0) node[pos=.5,scale=0.75] {$J_{2,2}$};

	\draw[pattern color=blue!60,pattern=north west lines] (2,0.0) rectangle (4,1.0) node[pos=.5,scale=0.75] {$J_{3,1}$};
	\draw[pattern color=blue!60,pattern=north west lines] (5,0.0) rectangle (6,1.0) node[pos=.5,scale=0.75] {$J_{3,1}$};

   \draw[->] (1.0,1.0) -- (1.0,2.20) node[pos=.5,scale=0.75,yshift=0.5cm] {$r_{3,1}$};

   \draw[->,red] (6,1.2) -- (6,0.0) node[pos=.5,scale=0.75,yshift=0.6cm] {$d_{3,1}$};     
    
    \end{tikzpicture}
    \caption{Stationary gang}
    \label{fig:strict_vs_stationary_stationary}
    \end{subfigure}
    \begin{subfigure}[b]{0.27\textwidth}
    \begin{tikzpicture}[xscale=0.5,yscale=0.5]

    \draw[->] (-0.1,-0.15) -- (8.5,-0.15);
\foreach \x in {0,...,8}
\draw (\x,-0.20) -- (\x,-0.1) node[yshift=-0.25cm,scale=0.6] {$\x$};

\draw[->] (-0.2,-0.15) -- (-0.2,2.3);
\draw (-0.3,0.5) -- (-0.1,0.5) node[xshift=-0.25cm,scale=0.6] {$P_2$};
\draw (-0.3,1.5) -- (-0.1,1.5) node[xshift=-0.25cm,scale=0.6] {$P_1$};

\draw[pattern color=black!60,pattern=north west lines] (0,0.0) rectangle (1.0,1.0) node[pos=.5,scale=0.75] {$J_{1,1}$};

\draw[pattern color=red!60,pattern=north west lines] (1.00,0.0) rectangle (2.00,2.0) node[pos=.5,scale=0.75] {$J_{2,1}$};

\draw[pattern color=black!60,pattern=north west lines] (3,0.0) rectangle (4.0,1.0) node[pos=.5,scale=0.75] {$J_{1,2}$};

\draw[pattern color=red!60,pattern=north west lines] (4,0.0) rectangle (5.0,2.0) node[pos=.5,scale=0.75] {$J_{2,2}$};

\draw[pattern color=blue!60,pattern=north west lines] (2,0.0) rectangle (3,1.0) node[pos=.5,scale=0.75] {$J_{3,1}$};

\draw[pattern color=blue!60,pattern=north west lines] (5,0.0) rectangle (6,1.0) node[pos=.5,scale=0.75] {$J_{3,1}$};

\draw[pattern color=black!60,pattern=north west lines] (6,0.0) rectangle (7.0,1.0) node[pos=.5,scale=0.75] {$J_{1,3}$};

\draw[pattern color=blue!60,pattern=north west lines] (7,0.0) rectangle (8,1.0) node[pos=.5,scale=0.75] {$J_{3,1}$};

   \draw[->] (0.0,1.0) -- (0.0,1.20) node[pos=.5,scale=0.75,yshift=0.15cm, xshift=0.30cm] {$r_{3,1}$};

\draw[->,red] (5.2,1.2) -- (5,1.0) -- (5,0.0) node[pos=.5,scale=0.75,yshift=0.58cm, xshift=0.44cm] {$d_{3,1}$};

    \end{tikzpicture}
    \caption{Strict partitioning}
    \label{fig:strict_vs_stationary_strict}
    \end{subfigure}
    \caption{Stationary scheduling better than strict partitioning.}
    \label{fig:strict_vs_stationary}
    \end{figure}
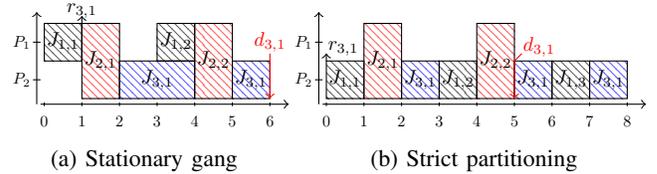 
\end{example}

\subsection{Global and Partitioned for Non-preemptive Gangs}
\label{sec:eg_np}

Non-preemptive scheduling of rigid gang tasks poses additional challenges.
It has been demonstrated that global policies for rigid gangs can lead to long priority-inversion~\cite{Dong:2019}.
To mitigate this issue, static partitioning can be employed to separate the tasks that cause priority inversion.

\begin{example}
\label{ex:2D}
Consider seven tasks $\tau_{1-7}$ that execute non-preemptively on three identical processors.
Task~$\tau_1$ must run on two processors simultaneously, while all other tasks can run on any single processor. Tasks worst-case execution times are as follows: $C_1{=}C_3{=}C_6{=}2$, $C_2{=}C_7{=}3$, and $C_4{=}4$.
A~priority inversion is when a higher-priority job must wait for the completion of a lower-priority job.
For example, all lower-priority tasks $\tau_{2-7}$ can be released one clock tick before~$\tau_1$.
A~global work-conserving scheduler always keeps the processors busy, and whenever a processor becomes idle, the next pending job that can run on idle processors is scheduled. 
While it results in good processor utilization, it can also lead to a  long priority inversion.
As shown in Figure~\ref{fig:np_global_vs_strict_global},
when the start and end times of old and new tasks overlap, the new tasks with smaller volumes can get ahead of pending big-volume tasks (\emph{i.e.,} $\tau_1$ cannot start as long as two processors are not~idle) and cause so-called~\emph{2D-blocking}~\cite{Dong:2019}.
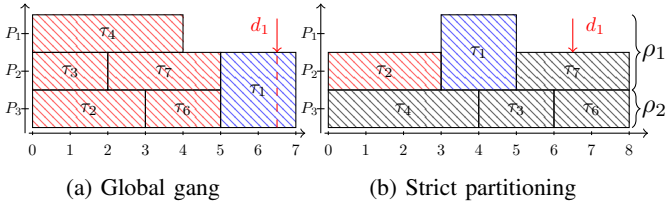
\begin{figure}[hbt]
\centering
\begin{subfigure}[b]{0.21\textwidth}
    \begin{tikzpicture}[xscale=0.5,yscale=0.5]
    
    \draw[->] (-0.1,-0.15) -- (7.25,-0.15);
    \foreach \x in {0,...,7}
        \draw (\x,-0.20) -- (\x,-0.1) node[yshift=-0.25cm,scale=0.6] {$\x$};

    \draw[->] (-0.2,-0.15) -- (-0.2,3.3);
    \draw (-0.3,0.5) -- (-0.1,0.5) node[xshift=-0.2cm,scale=0.6] {$P_3$};
    \draw (-0.3,1.5) -- (-0.1,1.5) node[xshift=-0.2cm,scale=0.6] {$P_2$};
    \draw (-0.3,2.5) -- (-0.1,2.5) node[xshift=-0.2cm,scale=0.6] {$P_1$};

	\draw[pattern color=red!60,pattern=north west lines] (0,0.0) rectangle (3.0,1.0) node[pos=.5,scale=0.75] {$\tau_2$};

	\draw[pattern color=red!60,pattern=north west lines] (3.00,0.0) rectangle (5.00,1.0) node[pos=.5,scale=0.75] {$\tau_{6}$};

    \draw[pattern color=red!60,pattern=north west lines] (0,2.0) rectangle (4.0,3.0) node[pos=.5,scale=0.75] {$\tau_{4}$};

    \draw[pattern color=red!60,pattern=north west lines] (0.0,1.0) rectangle (2.0,2.0) node[pos=.5,scale=0.75] {$\tau_{3}$};
    
    \draw[pattern color=red!60,pattern=north west lines] (2.0,1.0) rectangle (5.0,2.0) node[pos=.5,scale=0.75] {$\tau_{7}$};
    
	\draw[pattern color=blue!60,pattern=north west lines] (5,0.0) rectangle (7.0,2.0) node[pos=.5,scale=0.75] {$\tau_1$};

   \draw[<-,red] (6.5,2.0) -- (6.5,2.90) node[pos=.8,scale=0.75,xshift=-0.3cm] {$d_{1}$};  
   
   \draw[dashed,red] (6.5,1.9) -- (6.5,0) ;  
    
    \end{tikzpicture}
    \caption{Global gang}
    \label{fig:np_global_vs_strict_global}
    \end{subfigure}
    \begin{subfigure}[b]{0.26\textwidth}
    \begin{tikzpicture}[xscale=0.5,yscale=0.5]
	
	\draw[->] (-0.1,-0.15) -- (8.5,-0.15);
	\foreach \x in {0,...,8}
	\draw (\x,-0.20) -- (\x,-0.1) node[yshift=-0.25cm,scale=0.6] {$\x$};

	\draw[->] (-0.2,-0.15) -- (-0.2,3.3);
	\draw (-0.3,0.5) -- (-0.1,0.5) node[xshift=-0.2cm,scale=0.6] {$P_3$};
	\draw (-0.3,1.5) -- (-0.1,1.5) node[xshift=-0.2cm,scale=0.6] {$P_2$};
	\draw (-0.3,2.5) -- (-0.1,2.5) node[xshift=-0.2cm,scale=0.6] {$P_1$};

	\draw[pattern color=red!60,pattern=north west lines] (0,1.0) rectangle (3.0,2.0) node[pos=.5,scale=0.75] {$\tau_2$};
	
	\draw[pattern color=black!60,pattern=north west lines] (5.00,1.0) rectangle (8.00,2.0) node[pos=.5,scale=0.75] {$\tau_{7}$};
	
	\draw[pattern color=black!60,pattern=north west lines] (0,0.0) rectangle (4.0,1.0) node[pos=.5,scale=0.75] {$\tau_{4}$};

	\draw[pattern color=black!60,pattern=north west lines] (6.0,0.0) rectangle (8.0,1.0) node[pos=.5,scale=0.75] {$\tau_{6}$};
	
	\draw[pattern color=black!60,pattern=north west lines] (4.0,0.0) rectangle (6.0,1.0) node[pos=.5,scale=0.75] {$\tau_{3}$};
	
	\draw[pattern color=blue!60,pattern=north west lines] (3,1.0) rectangle (5.0,3.0) node[pos=.5,scale=0.75] {$\tau_1$};

	\draw[<-,red] (6.5,2.0) -- (6.5,2.90) node[pos=.8,scale=0.75,xshift=0.4cm] {$d_{1}$};  
	
	   \draw [decorate,decoration={brace,amplitude=3pt},xshift=5pt,yshift=0pt]  (7.9,3.0) -- (7.9,1.0) node [black,midway,xshift=0.3cm] {$\rho_1$}; 
	
	\draw [decorate,decoration={brace,amplitude=3pt},xshift=5pt,yshift=0pt]  (7.9,1.0) -- (7.9,0.0) node [black,midway,xshift=0.3cm] {$\rho_2$};

\end{tikzpicture}
    \caption{Strict partitioning}
    \label{fig:np_global_vs_strict_strict}
    \end{subfigure}
    \caption{Blocking in non-preemptive global and strict gang.}
    \label{fig:np_global_vs_strict}
    \end{figure}
\end{example}
Applying strict partitioning with a sequential scheduler on each partition makes it possible to find a compromise between parallelism and reduced blocking.
Figure~\ref{fig:np_global_vs_strict_strict} shows a strict partitioning with two partitions: $\rho_1=\{P_1, P_2\}$ and $\rho_2=\{P_3\}$.
While it leads to processor $P_1$ underutilization, the priority inversion experienced by $\tau_1$ can be reduced to the duration of the longest lower-priority task in $\rho_1$.

\section{Strict Partitioning Algorithms}
\label{sec:alg}

\subsection{Offline Partitioning Heuristic}

The strict partitioning strategy requires making two offline decisions: \textit{(i) how to divide processors into disjoint partitions} and \textit{(ii) how to assign each gang task to one of the processor partitions.} 
It is shown in~\cite{ekberg2021partitioned} that the task partitioning problem is NP-hard in the strong sense, even for sequential tasks.
Hence, the complexity extends to gang tasks.

We propose an offline strict partitioning heuristic \textit{First-fit Decreasing Volume (FFDV)} in Algorithm~\ref{alg:ffdv}, inspired by the first-fit decreasing-height algorithm for the 2-D \emph{strip packing} problem~\cite{coffman1980performance}. 
The heuristic first sorts the tasks by a non-increasing order of their volumes with ties broken by a non-decreasing order of task periods (line 2). Then, it assigns the sorted tasks one by one to form a sequence of partitions $\rho_1, \rho_2, ..., \rho_t$. Each task will be assigned to the first partition (\ie, with the lowest index) such that it is schedulable together with other tasks already assigned to this partition (lines 4 to~6). If none of the partitions can accommodate this task, a new partition will be created with the same volume as the task, given that there are sufficient processors left (lines 7 to 9). The algorithm terminates if no remaining tasks are pending to be assigned (\ie, return a success in line 12) or if there are no sufficient processors to create a new partition for the pending-to-be-assigned tasks (\ie, return a failure in line 11). 

We note that various schedulability tests can be used in line~4 to check the schedulability of the tasks assigned to each partition, depending on which online scheduler is used. We will present different variants combined with different types of online schedulers (\ie, uniprocessor task schedulers and global gang task schedulers) used for each partition in Section~\ref{sec:sp_algos}. 


  


{
\SetAlCapFnt{\small}
\SetAlCapNameFnt{\small}
\SetAlFnt{\small}
\begin{algorithm} [t]
    \caption{First Fit Decreasing Volume (FFDV)}
    \label{alg:ffdv}

    \KwIn{$\tau$: a gang task set to be scheduled;\newline $M$: number of available processors\;}
    \KwOut{$\rho, \tau(\rho)$: a set of partitions and the corresponding tasks assigned to each partition if schedulable; False otherwise.}
    $M' \gets M, \rho \gets \varnothing$\;
    \For{$\tau_i \in \tau$ (sorted by non-increasing order of volumes)}
    {
        \For{$\rho_j \in \rho$}
        {
            \If{$\tau(\rho_j) \cup \{\tau_i\}$ is schedulable on $\rho_j$}
            {
                $\tau(\rho_j) \gets \tau(\rho_j) \cup \{\tau_i\}$\;
                Continue to the next iteration in the outer loop\;
            }
        }
        \eIf{$m_i \leq M'$}
        {
            Create a partition $\rho_{new}$ with $\tau(\rho_{new}) = \{\tau_i\}$ and $|\rho_{new}| = m_i$\;
            $\rho \gets \rho \cup \{\rho_{new}\}, M' \gets M' - m_i$\;
        }
        {
            \Return False\;
        }
    }
    \Return $\rho$, $\tau(\rho) = \{\tau(\rho_j) ~|~ \rho_j \in \rho\}$\;
\end{algorithm}
}

The time complexity of the proposed FFDV heuristic is analyzed as follows.
There are two loops in Algorithm~\ref{alg:ffdv}. The outer loop iterates the task set, and the inner loop iterates the partitions that have been created and checks the schedulability. In the worst case, the numbers of tasks and partitions that need to be iterated are $n$ and $M$, respectively, leading to an overall time complexity of $\mathcal{O}(nM\Omega)$, where $\mathcal{O}(\Omega)$ denotes the time complexity of the schedulability test used in line 4.

\subsection{Two Strict Partitioning Variants: \texttt{SP-U} and \texttt{SP-G}}
\label{sec:sp_algos}

We propose two variants of the strict partitioning heuristic by considering two different types of online schedulers: \textit{\texttt{SP-U}} standing for the strict partitioning with \textit{uniprocessor online schedulers} and \textit{\texttt{SP-G}} standing for the strict partitioning with \textit{global gang online schedulers}.
The advantage of a global gang scheduler lies in its full utilization of processors by allowing different jobs to execute in parallel. In contrast, classical uniprocessor schedulers may lead to processor under-utilization, but they have lower runtime overhead and well-established exact schedulability tests. 
Both variants use FFDV as a framework to make offline partitioning decisions. 
For \texttt{SP-U}, we apply the FFDV heuristic as is by incorporating corresponding exact uniprocessor schedulability tests and prove its performance bounds in Section~\ref{sec:perf_bounds}.
For \texttt{SP-G}, we propose modifications to the FFDV heuristic to improve further its performance in Section~\ref{sec:spp}.

\subsection{Performance Bounds for \texttt{SP-U}}
\label{sec:perf_bounds}

We prove three performance bounds for the FFDV heuristic when using uniprocessor online schedulers (\ie, \texttt{SP-U}). 
The first bound (Theorem~\ref{the:wub}) is a weighted utilization bound defined by a weighting function widely used in bin-packing literature~\cite{garey1976resource}. The last two bounds (Theorem~\ref{the:gub_1} and \ref{the:gub}) are general utilization bounds inspired by the classical results in 2D strip packing literature~\cite{coffman1980performance}. 
In what follows, we denote by $u_b$ the utilization bound of a uniprocessor task scheduler (\eg, $u_b = 1$ for preemptive EDF~\cite{Liu:1973}).
\revise{Moreover, we denote by $FFDV(\tau)$ the number of processors used by the FFDV algorithm to schedule the task set $\tau$, and the resulting partitions as $\rho_1, \rho_2, ..., \rho_t$ (\ie, $t$ is the total number of partitions).} It suffices to prove the schedulability of $\tau$ under the FFDV algorithm on $M$ processors by showing $FFDV(\tau) \leq M$.

\begin{theorem}
\label{the:wub}
    A gang task set~$\tau$ is schedulable on $M$ processors by \texttt{SP-U} if 
    \begin{equation}
    \label{eq:wub}
        \sum_{\tau_i \in \tau}W(U_i) \leq (M - \overline{m}) \cdot u_b,
    \end{equation}
    where the weighting function $W: [0,u_b] \rightarrow [0,\frac{8}{5}u_b]$ is defined~as:
\begin{equation*}
    W(U_i) = 
    \begin{cases} 
    \frac{6}{5} U_i,  & \text{if }0 \leq U_i \leq \frac{1}{6} u_b; \\
    \frac{9}{5} U_i - \frac{1}{10},  & \text{if }\frac{1}{6} u_b < U_i \leq \frac{1}{3} u_b; \\
    \frac{6}{5} U_i + \frac{1}{10},  & \text{if }\frac{1}{3} u_b < U_i \leq \frac{1}{2} u_b; \\
    \frac{6}{5} U_i + \frac{4}{10},  & \text{if }\frac{1}{2} u_b < U_i \leq u_b.
    \end{cases}
\end{equation*}
\end{theorem}
\begin{IEEEproof}
We prove the theorem by showing that $FFDV(\tau) \leq M$ given inequality~\eqref{eq:wub}.
It is proved in Theorem~2 of \cite{coffman1980performance} that:
\begin{equation}
\label{eq:ffdv_wub}
    W(U) = \sum_{\tau_i \in \tau}W(U_i) \geq (FFDV(\tau) - \overline{m}) \cdot u_b.
\end{equation}
By combining~\eqref{eq:wub} and~\eqref{eq:ffdv_wub}, we have:
\begin{equation}
    (FFDV(\tau) - \overline{m}) \cdot u_b \leq \sum_{\tau_i \in \tau}W(U_i) \leq (M - \overline{m}) \cdot u_b.
\end{equation}
It follows that $FFDV(\tau) \leq M$.
\end{IEEEproof}

\begin{theorem}
\label{the:gub_1}
    A gang task set $\tau$ is schedulable on $M$ processors by \texttt{SP-U} if 
    \begin{equation}
    \label{eq:gub_1}
        U\leq \frac{(M - \overline{m} + \underline{m}) \cdot u_b}{2}.
    \end{equation}
\end{theorem}
\begin{IEEEproof}
    We denote the total utilization and sequential utilization of the tasks assigned to partition $\rho_i$ as $U(\rho_i)$ and $u(\rho_i)$, respectively. We also denote the sequential utilization of the first task assigned to $\rho_i$ as $u^0_i$. 
    For each $i \geq 2$, the first task in $\rho_i$ cannot fit into the previous partition $\rho_{i-1}$; therefore, we have $u(\rho_{i-1}) + u^0_i > u_b$. Since the tasks in $\rho_{i-1}$ have at least volume $|\rho_i|$, and the first task in $\rho_i$ has volume $|\rho_i|$, we have $U(\rho_{i-1}) + U(\rho_i) \geq (u(\rho_{i-1}) + u^0_i) |\rho_i| > |\rho_i| u_b$. Therefore,
    \begin{align*}
        FFDV(\tau) &= \sum_{i=1}^t{|\rho_i|} < |\rho_1| + \frac{1}{u_b} \left(\sum_{i=1}^{t-1}{U(\rho_i)} + \sum_{i=2}^{t}{U(\rho_i)}\right)\\
        &= \overline{m} + \frac{2}{u_b} \sum_{i=1}^t{U(\rho_i)} - \frac{1}{u_b} (U(\rho_1) + U(\rho_t)).
    \end{align*}
    Since $U(\rho_1) + U(\rho_t) > \underline{m} \cdot u_b$, it follows
    \begin{equation*}
        FFDV(\tau) < \overline{m} + \frac{2}{u_b} \sum_{i=1}^t{U(\rho_i)} - \underline{m}.
    \end{equation*}
    Since all the tasks are successfully assigned to $t$ partitions, we have $\sum_{i=1}^t{U(\rho_i)} = \sum_{\tau_i \in \tau}{U_i}=U$, thus $FFDV(\tau) < M$ follows given inequality~\eqref{eq:gub_1}. 
\end{IEEEproof}

\begin{theorem}
\label{the:gub}
    A gang task set $\tau$ is schedulable on $M$ processors by \texttt{SP-U} if $\exists p \in \mathbb{Z^+} \land p \geq 2$: 
    \begin{equation}
    \label{eq:max_util}
        u_i \leq \frac{u_b}{p}, \forall \tau_i \in \tau,
    \end{equation}
    and 
    \begin{equation}
    \label{eq:gub}
        U \leq \frac{p}{p+1} (M - \overline{m}) u_b.
    \end{equation}
\end{theorem}
\begin{IEEEproof}
We need to show $FFDV(\tau) \leq M$ given inequalities~\eqref{eq:max_util} and~\eqref{eq:gub}.
By the definition of the FFDV algorithm, we know $\rho_1 = \overline{m}$, and thus:
\begin{equation} 
\label{eq:gub_acc_m}
    FFDV(\tau) = \sum_{i=1}^{t}{|\rho_{i+1}|} + \overline{m},
\end{equation}
where $|\rho_{t+1}| = 0$. Therefore, it suffices to show that 
\begin{equation}
\label{eq:gub_suf_cond0}
    \frac{p}{p+1} \sum_{i=1}^{t}{|\rho_{i+1}|} u_b \leq 
    U,
\end{equation}
because \eqref{eq:gub} and \eqref{eq:gub_suf_cond0} $\implies \frac{p}{p+1} \sum_{i=1}^{t}{|\rho_{i+1}|} u_b \leq \frac{p}{p+1} (M - \overline{m}) u_b \implies \sum_{i=1}^{t}{|\rho_{i+1}|} + \overline{m} = FFDV(\tau) \leq M$.

The complete proof of inequality~\eqref{eq:gub_suf_cond0} is in Appendix A. 
\end{IEEEproof}

\subsection{Modifications of FFDV for \texttt{SP-G}}
\label{sec:spp}

We propose two modifications to the FFDV algorithm presented in Algorithm~\ref{alg:ffdv} based on two observations.

\begin{observation}
\label{obs:spp_test}
    \revise{Given a gang task set $\tau(\rho_j)$ assigned to a processor partition $\rho_j$, if 
    \begin{equation}
    \label{eq:seq_or_not}
        \forall \tau_x, \tau_y \in \tau(\rho_j): m_x + m_y > |\rho_j| ~(x \neq y),
    \end{equation}
    and $\tau(\rho_j)$ is deemed unschedulable on $\rho_j$ by an exact schedulability test for a uniprocessor task scheduler, $\tau(\rho_j)$ cannot be deemed schedulable on $\rho_j$ by any schedulability tests for a global gang task scheduler of the same type\footnote{For example, a global gang EDF scheduler is of the same type as a uniprocessor EDF scheduler.}. }
\end{observation}

The statement in Observation~\ref{obs:spp_test} is true since if \eqref{eq:seq_or_not} holds, any two jobs in $\tau(\rho_j)$ cannot run in parallel due to an insufficient number of processors in $\rho_j$. Therefore, the global gang task online scheduler degrades to a uniprocessor task scheduler of the same type, and the exact analysis for the uniprocessor scheduler is thus an optimal schedulability test for such task sets. 

Inspired by the above observation, we propose checking the condition in Inequality~\eqref{eq:seq_or_not} before running the schedulability test. If the condition is satisfied (\ie, no tasks can run in parallel), we will use an exact analysis for the uniprocessor task scheduler as our schedulability test. Otherwise, we will use the sufficient (but not necessary\footnote{To the best of our knowledge, there has been no exact schedulability analysis for global gang task schedulers.}) schedulability analysis for the global gang task scheduler. Specific procedures of this modification are presented in Algorithm~\ref{alg:spp_test}, where \texttt{uniTest} and \texttt{globalTest} denote the exact analysis for uniprocessor scheduler and the sufficient analysis for global gang task scheduler, respectively. 

{
\SetAlCapFnt{\small}
\SetAlCapNameFnt{\small}
\SetAlFnt{\small}
\begin{algorithm} [t]
    \caption{Schedulability Test Procedures for \texttt{SP-P} (to replace lines 4 to 6 of Algorithm~\ref{alg:ffdv})}
    \label{alg:spp_test}

            \eIf{$\forall \tau_x, \tau_y \in \tau(\rho_j) \cup \{\tau_i\} : m_x + m_y > |\rho_j| ~(x \neq y)$}
            {
                \texttt{Test} $\gets$ \texttt{uniTest}\;
            }
            {
                \texttt{Test} $\gets$ \texttt{globalTest}\;
            }
            \If{$\tau(\rho_j) \cup \{\tau_i\}$ is schedulable on $\rho_j$ according to \texttt{Test}}
            {
                $\tau(\rho_j) \gets \tau(\rho_j) \cup \{\tau_i\}$\;
                Continue to the next iteration in the outer loop\;
            }
            
\end{algorithm}
}

The second observation is that we can leverage the unused processors to increase the volume of an existing partition when a task cannot be scheduled on any existing partitions and the remaining processors are insufficient to create a new partition with the same volume as the unschedulable task. Here, we propose to increase the volume of the last partition in the partition list since the task volumes on the last partition are the closest to those of the tasks that are yet to be assigned. 

\begin{observation}
    In \texttt{SP-G}, if a task $\tau_i$ cannot be assigned to any existing partitions $\rho_1, \rho_2, ..., \rho_t$ and the number of remaining processors $M'$ satisfies $0 < M' < m_i$, there is a possibility for $\tau_i$ to be schedulable on the last partition $\rho_t$ by increasing its volume by $M'$.
\end{observation}

Based on this observation, we introduce the volume increase procedures in Algorithm~\ref{alg:spp_volume} to replace lines 10 to 11 of Algorithm~\ref{alg:ffdv} for improving the FFDV performance.
We note that this modification is not necessarily applied to the \texttt{SP-U} variant since increasing the volume of existing partitions cannot make unschedulable tasks become schedulable due to the nature of uniprocessor task schedulers (\ie, all the processors in a partition are regarded as a single resource). 

{
\SetAlCapFnt{\small}
\SetAlCapNameFnt{\small}
\SetAlFnt{\small}
\begin{algorithm} [t]
    \caption{Volume Increase Procedures for \texttt{SP-P} (to replace lines 10 to 11 of Algorithm~\ref{alg:ffdv})}
    \label{alg:spp_volume}

        \ElseIf{$M' > 0$ and $\tau(\rho_j) \cup \{\tau_i\}$ is schedulable with volume $|\rho_j| + M'$}
        {
            Increase $\rho_j$'s volume to $|\rho_j| + M'$\;
            $\tau(\rho_j) \gets \tau(\rho_j) \cup \{\tau_i\}, M' \gets 0$\;
            Continue to the next iteration in the outer loop\;
        }
        \Else
        {
            \Return False\;
        }
\end{algorithm}
}

\section{Performance Evaluation}
\label{sec:exp}

\begin{figure*}[ht]
    \centering
    \begin{subfigure}{.33\textwidth}
      \centering
      \includegraphics[width=0.95\linewidth]{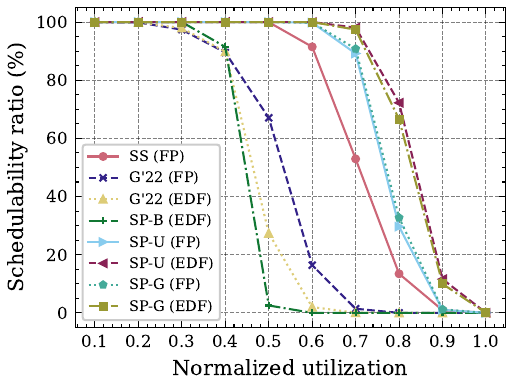}
      \caption{Low volume level}
      \label{fig:fp_s}
    \end{subfigure}%
    \begin{subfigure}{.33\textwidth}
      \centering
      \includegraphics[width=0.95\linewidth]{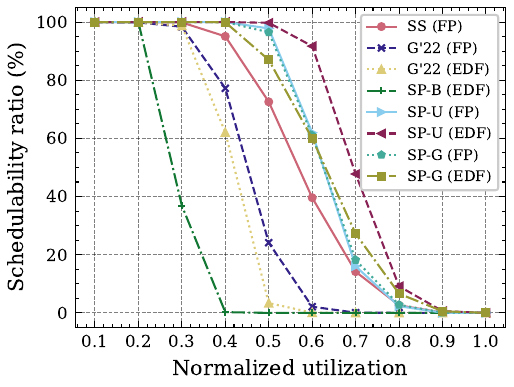}
      \caption{Medium volume level}
      \label{fig:fp_m}
    \end{subfigure}%
    \begin{subfigure}{.33\textwidth}
      \centering
      \includegraphics[width=0.95\linewidth]{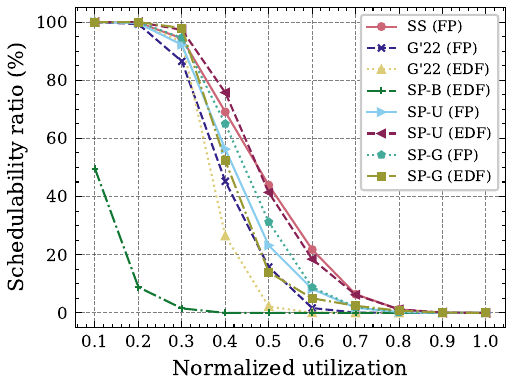}
      \caption{High volume level}
      \label{fig:fp_l}
    \end{subfigure}%
    \caption{Schedulability ratio (\textit{i.e.}, $\frac{\text{\#schedulable task sets}}{\text{\#total task sets}} \times 100$ \%) of preemptive gang task sets.}
    \label{fig:res_fp}
\end{figure*}

\begin{figure*}[ht]
    \centering
    \begin{subfigure}{.33\textwidth}
      \centering
      \includegraphics[width=0.95\linewidth]{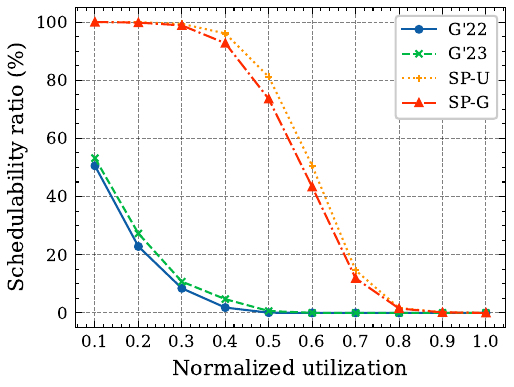}
      \caption{Low volume level}
      \label{fig:np_s}
    \end{subfigure}%
    \begin{subfigure}{.33\textwidth}
      \centering
      \includegraphics[width=0.95\linewidth]{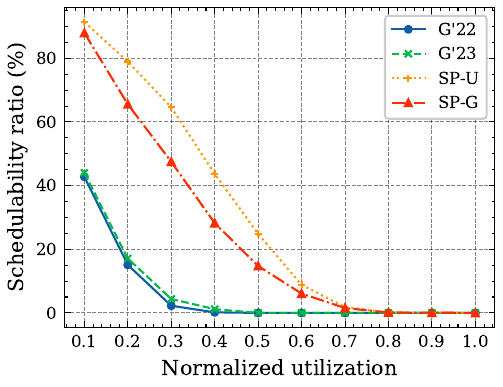}
      \caption{Medium volume level}
      \label{fig:np_m}
    \end{subfigure}%
    \begin{subfigure}{.33\textwidth}
      \centering
      \includegraphics[width=0.95\linewidth]{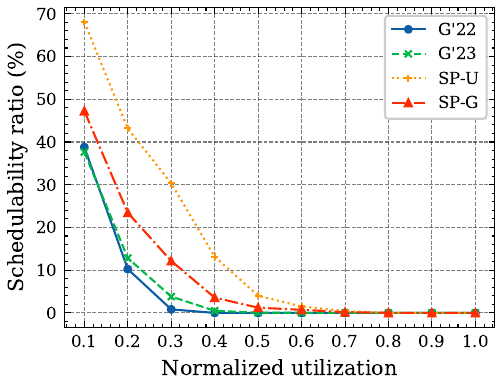}
      \caption{High volume level}
      \label{fig:np_l}
    \end{subfigure}%
    \caption{Schedulability ratio (\textit{i.e.}, $\frac{\text{\#schedulable task sets}}{\text{\#total task sets}} \times 100$ \%) of non-preemptive gang task sets.}
    \label{fig:res_np}
\end{figure*}

This section evaluates the schedulability performance of the proposed strict partitioning strategy by comparing it with state-of-the-art schedulability analyses for gang task sets. 
We perform two sets of experiments. First, from Section \ref{sec:synthetic_setting} to \ref{sec:np_res}, we experiment with comprehensive synthetic task sets generated by standard task set generation tools. Second, in Section \ref{sec:case_study}, we conduct a case study using Edge TPU benchmarks of representative DNNs widely used in computer vision applications. For synthetic task sets, we consider both preemptive and non-preemptive task sets with both FP and EDF scheduling policies. For the Edge TPU case study, neural network inference tasks are modeled as non-preemptive gangs due to the software and hardware characteristics of Edge TPU.

\subsection{Synthetic Evaluation Settings}
\label{sec:synthetic_setting}

\vspace{0.5em}
\noindent
\textit{Task set generation.}
We generate synthetic task sets based on a standard task set generation tool \texttt{DRS} \cite{griffin2020generating}. For a systematic evaluation, we vary four task set parameters:
(i) number of processors $M \in \{8, 16\}$; (ii) number of tasks $n \in \{M, 2M\}$; (iii) task volume $m_i$ range $\in \{[1, \lceil0.3M\rceil]$ (low volume), $[1, \lceil0.6M\rceil]$ (medium volume), $[1, M)$ (large volume)$\}$; (iv) normalized task set utilization $U / M \in \{0.1, 0.2, ..., 1.0\}$.

For each parameter combination, we generate 1,000 task sets. The generation of each task set includes the following procedures. First, we use \texttt{DRS} to generate task utilization $U_i$ such that the sum equals the target task set utilization ($U = \sum_{\tau_i \in \tau}U_i$). Note that we set the maximum utilization for each task to be not greater than the task volume upper bound to avoid infeasible sequential task utilization (\ie, $u_i > 1$). Second, we uniformly sample task periods $T_i$ and volumes $m_i$ from $[10, 1000]$ ms and the target volume range, respectively. Similarly, we set the minimum volume as the ceil of task utilization (\ie, $\lceil U_i \rceil$) to avoid infeasible sequential utilization. Finally, we calculate the task WCETs by $C_i = \lfloor U_i \cdot T_i / m_i \rfloor$.

\vspace{0.5em}
\noindent
\textit{Comparison Algorithms.}
We compare the schedulability ratio of both preemptive and non-preemptive scheduling techniques on the same generated task sets. For preemptive scheduling, we consider both FP and EDF schedulers and compare the proposed strict partitioning with the following state-of-the-art schedulability analyses for gang tasks. 
\begin{itemize}
    \item \texttt{SS} (preemptive FP): stationary scheduling for preemptive FP in \cite{Ueter:2021}. 
    \item \texttt{G'22} (preemptive FP/EDF): global schedulability analyses for preemptive FP/EDF in \cite{lee2022response}.
    \item \revise{\texttt{SP-B} (preemptive EDF): strict partitioning utilization bound for preemptive EDF (combination of Theorem~\ref{the:wub}, \ref{the:gub_1}, and \ref{the:gub}, \ie, a task set is deemed schedulable if any of the utilization bound conditions is satisfied).} 
    \item \texttt{SP-U} (preemptive FP/EDF): strict partitioning with uniprocessor online schedulers for preemptive FP/EDF. The exact preemptive FP response time analysis~\cite{Audsley:1993} and EDF utilization bound~\cite{Liu:1973} are used for checking the schedulability.
    \item \texttt{SP-G} (preemptive FP/EDF): strict partitioning with global gang online schedulers for preemptive FP/EDF. We use \texttt{G'22} as the \texttt{globalTest} in Algorithm~\ref{alg:spp_test}.
\end{itemize}

\revise{
For non-preemptive scheduling, we focus our comparison on FP policies.
The comparison algorithms include:}
\begin{itemize}
    \item \texttt{G'22} (non-preemptive FP): global schedulability analysis for non-preemptive FP in \cite{Lee:2022}.
    \item \texttt{G'23} (non-preemptive FP): global schedulability analysis for non-preemptive FP in \cite{Binqi:2023}.
    \item \texttt{SP-U} (non-preemptive FP): strict partitioning with uniprocessor online schedulers for non-preemptive FP. The exact non-preemptive FP response time analysis in~\cite{Davis:2007} is used for checking the schedulability.
    \item \texttt{SP-G} (non-preemptive FP): strict partitioning with global gang online schedulers for non-preemptive FP. We use \texttt{G'23} as the \texttt{globalTest} in Algorithm~\ref{alg:spp_test}.
\end{itemize}

For a fair comparison of all schedulability tests with FP policy, we apply DM~\cite{Audsley:1991} as the priority assignment algorithm. 
For each schedulability test, we calculate its schedulability ratio and present the results in Figure \ref{fig:res_fp} and Figure \ref{fig:res_np} for the preemptive and the non-preemptive cases, respectively\footnote{The figures show the average schedulability ratio over different $M$ and $n$.}.

\subsection{Evaluation Results of Preemptive Gang Scheduling}
The evaluation results of preemptive rigid gang scheduling are shown in Figure~\ref{fig:res_fp}, supporting the following observations. 

\vspace{0.5em}
\noindent
\textit{Global scheduling vs partitioned scheduling.}
In all scenarios, partitioned scheduling (\ie, \texttt{SS}, \texttt{SP-U}, and \texttt{SP-G}) achieved higher schedulability ratio than the state-of-the-art global schedulability analysis (\ie, \texttt{G'22}).
\revise{Specifically, for FP, the best performing partitioned scheduling method achieved up to 89.4\%, 73.7\%, and 28.2\% more schedulable task sets than \texttt{G'22}, respectively. For EDF, the better performing strict partitioning method achieved up to 98.0\%, 96.4\%, and 49.3\% more schedulable task sets than \texttt{G'22}, respectively.} 
The results verify that the schedulability tests for partitioned scheduling are less pessimistic than the global analyses (as discussed in Section~\ref{sec:eg_global}), and this advantage compensates for the loss due to processor under-utilization. 
\revise{Moreover, the utilization bound test \texttt{SP-B} performs better than \texttt{G'22} (EDF) for low-volume and low-utilization task sets. We note that the utilization bound test has linear time complexity.}

\vspace{0.5em}
\noindent
\textit{Stationary scheduling vs strict partitioning.}
For all scenarios with \textit{low or medium} task volume levels, all strict partitioning variants (\ie, \texttt{SP-U} and \texttt{SP-G} with FP/EDF) achieved a higher schedulability ratio than stationary scheduling. 
This observation demonstrates the effectiveness of avoiding inter-partition interference in strict partitioning, as discussed in Section~\ref{sec:eg_stationary}.
For scenarios with \textit{large} task volume levels, stationary scheduling achieved a higher schedulability ratio than the strict partitioning variants except for \texttt{SP-U} (EDF). 
This is expected since, for task sets with high-volume tasks, very few partitions can be created in strict partitioning. As a result, the interference between tasks cannot be effectively reduced by building boundaries through disjoint partitions.

\vspace{0.5em}
\noindent
\textit{\texttt{SP-U} vs \texttt{SP-G}.} 
For scenarios with low task volume levels, \texttt{SP-U} and \texttt{SP-G} achieved similar schedulability ratios. 
This is expected since, for task sets with low task volume levels, there is a high chance that the tasks assigned to the same partition have a similar volume to the volume of the partition. As a result, tasks cannot execute in parallel on most partitions, and thus \texttt{SP-G} will use the same schedulability tests as \texttt{SP-U} according to Algorithm~\ref{alg:spp_test}.
For scenarios with high task volume levels, \texttt{SP-G} achieved a higher schedulability ratio than \texttt{SP-U} \revise{by up to 8.7\%} for the FP scheduling policy while a lower schedulability ratio \revise{by up to 27.4\%} for the EDF. 
This demonstrates that, compared to their corresponding uniprocessor exact analyses, there is less pessimism in global FP gang analysis than in global EDF.

\vspace{0.5em}
\noindent
\textit{FP vs EDF.}
The global gang schedulability tests (\ie, \texttt{G'22}) achieved a higher schedulability ratio for FP than EDF, while \texttt{SP-U} achieved a higher schedulability ratio for EDF than FP. 
This observation reveals that the uniprocessor preemptive EDF test achieves higher schedulability than FP, which is a well-known result. However, the results of the corresponding global gang analyses are exactly the opposite. 
For \texttt{SP-G}, it achieved a higher schedulability ratio for EDF than FP with low-volume tasks, while a higher schedulability ratio for FP than EDF with high-volume tasks. 
This is reasonable since, for low-volume tasks, \texttt{SP-G} benefits from using uniprocessor exact tests on most partitions. In contrast, for high-volume tasks, \texttt{SP-G} uses global analysis more often since tasks are more likely to execute in parallel.

\subsection{Evaluation Results of Non-preemptive Gang Scheduling}
\label{sec:np_res}

The schedulability results of non-preemptive gang scheduling are shown in Figure~\ref{fig:res_np}.

\vspace{0.5em}
\noindent
\textit{Global scheduling vs strict partitioning.}
In all scenarios, both strict partitioning variants (\ie, \texttt{SP-U} and \texttt{SP-G}) achieved a higher schedulability ratio than global scheduling analyses. 
\revise{Specifically, for low-, medium-, and high-volume task sets, the bettering performing strict partitioning method achieved up to 91.4\%, 62.0\%, and 30.4\%, respectively. }
This demonstrates the effectiveness of strict partitioning in reducing the pessimism of schedulability analyses (\eg, 2D-blocking~\cite{Dong:2019}) by assigning tasks with similar volumes to the same partitions and avoiding inter-partition interference, as discussed in Section~\ref{sec:eg_np}.

\vspace{0.5em}
\noindent
\textit{\texttt{SP-U} vs \texttt{SP-G}.} 
For scenarios with a low task volume level, \texttt{SP-U} and \texttt{SP-G} achieved similar schedulability ratios. For scenarios with a \textit{medium or large} task volume level, \texttt{SP-U} achieved a higher schedulability ratio than \texttt{SP-G}, and the advantage increases with the task volume level. The first observation is explained in the same way as the preemptive case. The second observation reveals that the non-preemptive global gang analyses are so pessimistic that it is better to use the uniprocessor exact test even when tasks can run in parallel.

\subsection{Edge TPU Case Study}
\label{sec:case_study}

We perform a case study based on the benchmarks of DNN workloads in computer vision applications on an AI hardware accelerator (\ie, Edge TPU) to evaluate the performance of the proposed strict partitioning strategy on real systems. 

\subsubsection{Hardware and Software Specifications}
\begin{figure}[t]
\centering
\includegraphics[width=0.9\linewidth]{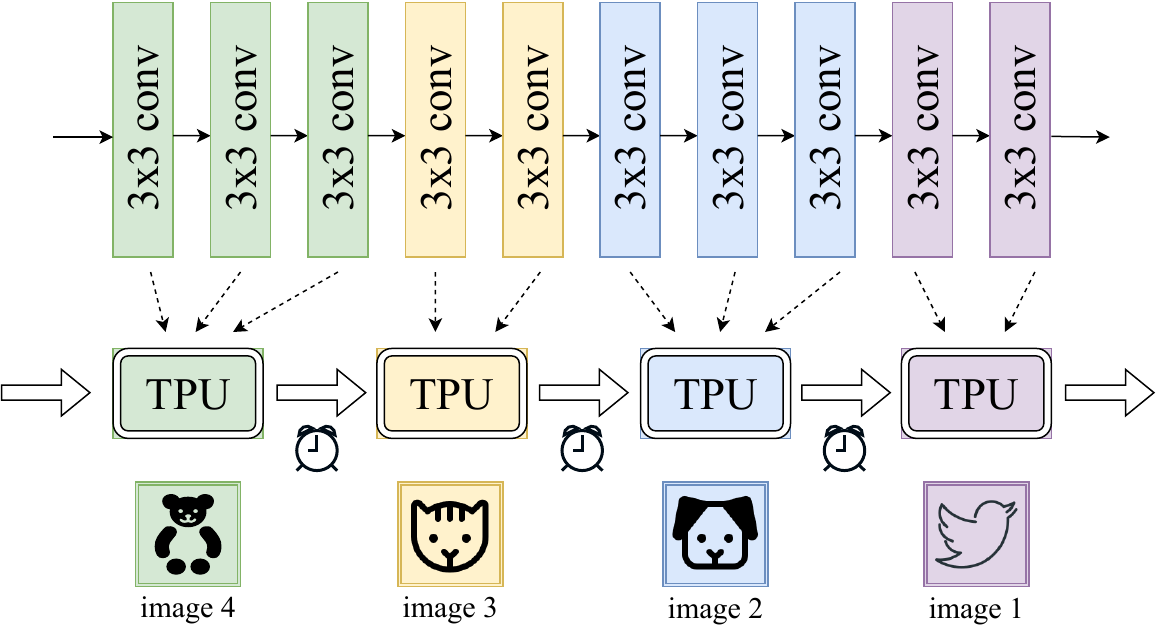}
\caption{Inllustration of Edge TPU pipelining.}
\label{fig:pipeline}
\end{figure}

In our case study, we use the COTS ASUS AI Accelerator cards CRL-G18U-P3D and CRL-G116U-P3D\footnote{https://iot.asus.com/products/AI-accelerator/AI-Accelerator-PCIe-Card/}, integrated with 8 and 16 Edge TPUs, respectively. Each Edge TPU has an on-chip scratchpad memory of 8 MB, which can be used for storing DNN weights. If the weights of a DNN model are larger than 8 MB, some weights cannot be fit into the on-chip memory and thus need to be loaded from the external memory (\ie,~DRAM) at runtime. 
This will incur a large runtime overhead (see Figure~\ref{fig:config_time}) and increase DRAM bandwidth contention.
To avoid these issues, we can use the \textit{Edge TPU Pipelining} technique. It splits a large DNN model into several consecutive segments and assigns each segment onto one Edge TPU such that the weights of each segment do not exceed the on-chip memory limit. This way, we can use multiple Edge TPU segments to form a pipeline. Figure~\ref{fig:pipeline} illustrates the Edge TPU pipelining technique, where four images are processed on four different Edge TPU segments in a pipeline, and we use different colors to show the mapping between DNN and Edge TPU segments.

\begin{figure}[t]
\centering
\includegraphics[width=0.8\linewidth]{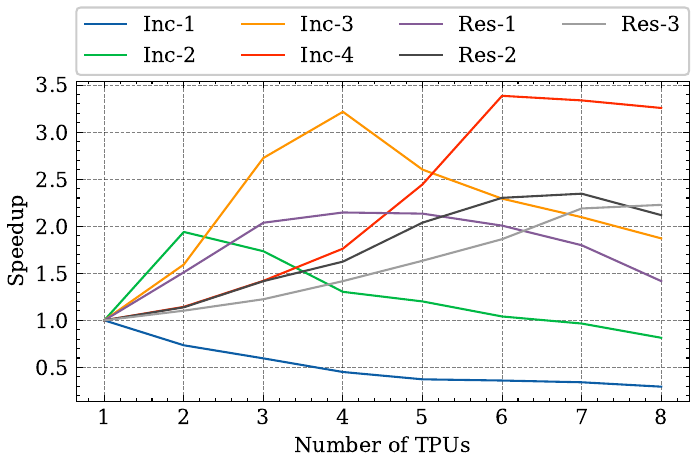}
\caption{Speedup factors of benchmarked DNN models w.r.t. different number of Edge TPUs on CRL-G18U-P3D.}
\vspace{-1em}
\label{fig:bench}
\end{figure}

There are two main advantages of using the Edge TPU pipelining technique. First, it avoids the aforementioned issues of loading weights from DRAM at runtime. Second, it enables pipeline parallelism since different inputs can be processed on different Edge TPU segments at the same time. The downside is the additional overhead caused by the communication time between two consecutive Edge TPUs.
Therefore, small models benefit less from running on a large number of Edge TPUs than large models, considering the tradeoff between communication overhead and on-chip caching.
Figure~\ref{fig:bench} shows the tradeoff by presenting the speedup factors of DNNs with different sizes running on Edge TPU pipelines with different volumes.

Once we decide the number of Edge TPUs to be used for a DNN model, we compile the DNN model into executable files using the Edge TPU Compiler\footnote{\url{https://coral.ai/docs/edgetpu/compiler/}} and execute the model using the \texttt{libcoral} runtime library\footnote{\url{https://coral.ai/docs/reference/cpp/}}. The \texttt{libcoral} runtime spawns one thread for each Edge TPU segment and allocates the on-chip memory in all threads simultaneously. After the DNN inferences are finished and the results are retrieved, all the threads will be terminated at the same time. Since a DNN model occupies multiple Edge TPU segments, with all the threads starting and finishing synchronously, the DNN inference on Edge TPUs can be modeled as a gang task. Moreover, the execution is non-preemptive since the \texttt{libcoral} runtime does not support preemption within a DNN inference.

\subsubsection{Workloads}
We benchmarked seven representative DNN models widely used in computer vision applications. For each DNN model, we compile it with all possible numbers of Edge TPUs on the card. For each Edge TPU number, we measure its corresponding worst observed execution time (\emph{WOET}) by running the model 1,000 times with different input images. Then, we set the task volume as the Edge TPU number that leads to the minimum worst observed execution time. Table~\ref{tab:tpu_bench} summarizes the specifications of the benchmarked workloads. 

After we get the Edge TPU benchmarks, we generate two test suites for evaluation. The first test suite consists of the first six DNNs shown in Table~\ref{tab:tpu_bench} and considers CRL-G18U-P3D with 8 Edge TPUs as the hardware platform, while the second suite contains all the benchmarked DNNs and uses CRL-G116U-P3D with 16 Edge TPUs since the last DNN ResNet-152 requires a volume of 9. 
For each test suite, we vary the normalized task set utilization from 0.1 to 1.0 with a step of 0.1, and generate random task utilization for each DNN workload using \texttt{DRS}. The task periods are calculated by $T_i = C_i \cdot m_i / U_i$ accordingly. 

\begin{table}[t] 
  \begin{center}
  \scriptsize
    \caption{Edge TPU benchmarks}
    \label{tab:tpu_bench}
    \begin{tabular}{|l|c|c|c|c|c|c|c|}
    \hline
    \textbf{Model} & Inc-1 & Inc-2 & Inc-3 & Inc-4 & Res-1 & Res-2 & Res-3 \\ \hhline{|=|=|=|=|=|=|=|=|}
    \textbf{Size (MB)} & 5.72 & 10.19 & 21.56 & 40.90 & 23.40 & 42.46 & 57.53 \\ \hline
    \textbf{Volume} & 1 & 2 & 4 & 6 & 4 & 7 & 9 \\ \hline
    \textbf{WOET (ms)} & $6$ & $10$ & $15$ & $31$ & $24$ & $44$ & $55$ \\ \hline
    \end{tabular}
  \end{center}
\end{table}

We generate 1,000 task sets for each target normalized task set utilization and compare the schedulability ratio of different schedulability tests. Since the DNN tasks are modeled as non-preemptive gang tasks, only non-preemptive gang schedulability tests are used for evaluation.

\subsubsection{Evaluation Results}
\begin{figure}[ht]
    \centering
    \begin{subfigure}{.5\textwidth}
      \centering
      \includegraphics[width=0.8\linewidth]{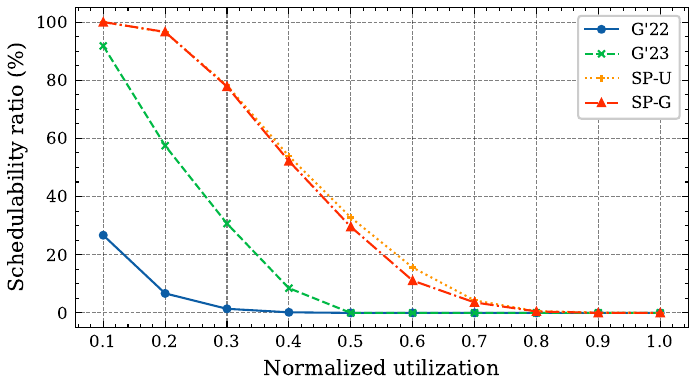}
      \caption{$n=6, M=8$}
      \label{fig:tpu_n6_m8}
    \end{subfigure}%
    \\
    \vspace{0.5em}
    \begin{subfigure}{.5\textwidth}
      \centering
      \includegraphics[width=0.8\linewidth]{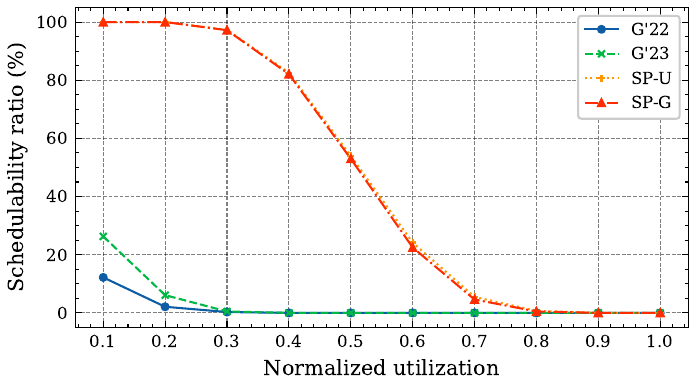}
      \caption{$n=7, M=16$}
      \label{fig:tpu_n7_m16}
    \end{subfigure}%
    \caption{Schedulability ratio (\textit{i.e.}, $\frac{\text{\#schedulable task sets}}{\text{\#total task sets}} \times 100$ \%) of non-preemptive gang DNN task sets on Edge TPU pipelines.}
    \label{fig:res_tpu}
\end{figure}

Figure~\ref{fig:res_tpu} depicts the schedulability ratio of different schedulability tests. The observations are similar to those of the synthetic non-preemptive task sets shown in Figure~\ref{fig:res_np}, \ie, the strict partitioning strategy outperforms global non-preemptive gang analyses, and \texttt{SP-U} achieves similar schedulability ratio to \texttt{SP-G} for relatively low volume levels (Figure~\ref{fig:tpu_n7_m16}). It demonstrates the effectiveness of the proposed strict partitioning strategy on real systems and the potential of using a classical uniprocessor scheduler for each partition to achieve low overhead and high schedulability performance.

\section{Conclusion}
\label{sec:conclusion}

This paper proposed a new partitioned scheduling strategy, named \textit{strict partitioning}, for gang tasks. Its main idea is to divide tasks and processors into disjoint subsets and assign tasks statically onto processor partitions. Strict partitioning has four main advantages. 
(i) It builds boundaries between different partitions so there is no inter-partition interference. 
(ii) It tries to assign tasks with similar volumes to the same partition so that the intra-partition interference is reduced. 
(iii)~The tasks within each partition can be scheduled by any type of scheduler, which allows the use of a tighter schedulability test. 
(iv) The tasks are statically assigned to processor partitions so that there is less overhead caused by task migration. 
Experiments on synthetic task sets and a case study based on Edge TPU benchmarks demonstrated the effectiveness of the proposed strategy and algorithms by comparing them with state-of-the-art gang scheduling techniques. 

In future works, we plan to study the hybrid of strict partitioning and global sequential task scheduling~\cite{Marko:2007,Marko:2009,Baker:2003} for rigid gang tasks. Semi-partitioned gang scheduling is also an interesting strategy to research.

\appendices
\section{Proof of Inequality~\eqref{eq:gub_suf_cond0} in Theorem~\ref{the:gub}}

Let us define $U_{ij}$ and $u_{ij}$ as the total task utilization and sequential task utilization, respectively, of the tasks assigned to partition $\rho_j$ when the last task is assigned in $\rho_i$ before the next partition $\rho_{i+1}$ is created. Moreover, we define $u^0_i$ as the sequential utilization of the first task assigned to partition $\rho_i$ and $\delta_i = \max(0, p/(p+1)u_b - u_{ii})$.
Before proving inequality~\eqref{eq:gub_suf_cond0}, we first present the following lemmas that are essential to our proof.

\begin{lemma}
\label{lem:1}
    The following inequality holds.
    \begin{equation}
    \label{eq:U_ii}
        U_{ii} \geq u_{ii} |\rho_{i+1}| + u^0_i (|\rho_{i}| - |\rho_{i+1}|).
    \end{equation}
\end{lemma}
\revise{
\begin{IEEEproof}
We denote by $\tau^0_i$ the first task assigned to partition $\rho_i$ and by $\tau(\rho_i)$ the set of tasks assigned to $\rho_i$ when the last task is assigned to $\rho_i$ before the next partition $\rho_{i+1}$ is created. 
By the definition of $U_{ii}$, the LHS of \eqref{eq:U_ii} is the total utilization of $\tau(\rho_i)$. 
The RHS of \eqref{eq:U_ii} $= u^0_i |\rho_{i}| + (u_{ii} - u^0_i) |\rho_{i+1}|$. Since $u^0_i$ and $|\rho_{i}|$ are the sequential utilization and the volume of $\tau^0_i$, respectively, $u^0_i |\rho_{i}|$ is the utilization of $\tau^0_i$ (denoted as $U(\tau^0_i)$). Moreover, by the definition of $u_{ii}$, $(u_{ii} - u^0_i)$ is the total sequential utilization of $\tau(\rho_i) \setminus \{\tau^0_i\}$. Since the tasks are assigned in the decreasing order of volumes, the volumes of the tasks in $\tau(\rho_i) \setminus \{\tau^0_i\}$ are not smaller than $|\rho_{i+1}|$. Therefore, the total utilization of $\tau(\rho_i) \setminus \{\tau^0_i\}$ (denoted as $U (\tau(\rho_i) \setminus \{\tau^0_i\})$) is not smaller than $(u_{ii} - u^0_i) |\rho_{i+1}|$. 
Thus, $U_{ii} = U(\tau^0_i) + U (\tau(\rho_i) \setminus \{\tau^0_i\}) \geq u^0_i |\rho_{i}| + (u_{ii} - u^0_i) |\rho_{i+1}|$.
\end{IEEEproof}

\begin{lemma}
\label{lem:2}
    The following inequality holds.
    \begin{equation}
    \label{eq:u_ii}
        u_{ii} \geq \left(\frac{p}{p+1} u_b - \delta_i\right).
    \end{equation}
\end{lemma}
\begin{IEEEproof}
    For \eqref{eq:u_ii}, there are two cases to consider: $\delta_i = 0$ and $\delta_i > 0$. If $\delta_i = 0$, by the definition of $\delta_i$, we know that $u_{ii} \geq p/(p+1) u_b$. Thus, \eqref{eq:u_ii} holds. If $\delta_i > 0$, we know $\delta_i = p/(p+1) u_b - u_{ii}$, \eqref{eq:u_ii} still holds.
\end{IEEEproof}

\begin{lemma}
\label{lem:3}
    The following inequality holds.
    \begin{equation}
    \label{eq:U_ii_2}
        U_{ii} \geq \left(\frac{p}{p+1} u_b - \delta_i\right) |\rho_{i+1}|.
    \end{equation}
\end{lemma}
\begin{IEEEproof}
    Take \eqref{eq:u_ii} into \eqref{eq:U_ii}, we have $U_{ii} \geq \left(\frac{p}{p+1} u_b - \delta_i\right) |\rho_{i+1}| + u^0_i (|\rho_{i}| - |\rho_{i+1}|)$. Since $|\rho_{i}| \geq |\rho_{i+1}|$ and $u^0_i \geq 0$, $U_{ii} \geq \left(\frac{p}{p+1} u_b - \delta_i\right) |\rho_{i+1}|$. 
\end{IEEEproof}

\begin{lemma}
\label{lem:4}
    If $\delta_{i-1} > 0$, the following inequality holds.
    \begin{equation}
    \label{eq:u_0_2}
        u^0_i \geq \frac{1}{p+1}u_b + \delta_{i-1}.
    \end{equation}
\end{lemma}
\begin{IEEEproof}
    Since $\delta_{i-1} > 0$,
    $\delta_{i-1} = p / (p+1) u_b - u_{i-1,i-1} > 0$.
    Moreover, since $u^0_i \geq u_b - u_{i-1,i-1}$,
    \eqref{eq:u_0_2} follows.
\end{IEEEproof}

Now, we prove \eqref{eq:gub_suf_cond0} based on Lemmas~\ref{lem:1}-\ref{lem:4}.
}

\smallskip
\noindent
\textbf{Proof of Inequality~\eqref{eq:gub_suf_cond0} in Theorem~\ref{the:gub}.}

Since all the tasks in $\tau$ are successfully assigned by the FFDV algorithm to partitions $\rho_1, \rho_2, ..., \rho_t$, we know $U = \sum_{j=1}^tU_{tj}$.
Replace $U$ by $\sum_{j=1}^tU_{tj}$ in~\eqref{eq:gub_suf_cond0}, it suffices to prove:
\begin{equation}
\label{eq:gub_suf_cond}
    \sum_{i=1}^tU_{tj} \geq \frac{p}{p+1} \sum_{i=1}^{t}{|\rho_{i+1}|} u_b.
\end{equation}
We prove~\eqref{eq:gub_suf_cond} by induction. 
In what follows, we will prove
\begin{equation}
\label{eq:gub_induct}
    \forall i \in [1,t]: \sum_{j=1}^iU_{ij} \geq \frac{p}{p+1} \sum_{j=1}^i{|\rho_{j+1}|} u_b - \delta_i |\rho_{i+1}|,
\end{equation}
which implies~\eqref{eq:gub_suf_cond} by taking $i=t$.

For $i=1$, \eqref{eq:gub_induct} is reduced to 
\begin{equation*}
\label{eq:U_11}
    U_{11} \geq \left(\frac{p}{p+1} u_b - \delta_1\right) |\rho_2|,
\end{equation*}
which is proved by setting $i=1$ in \eqref{eq:U_ii_2}. 

For $1 < i \leq t$, we prove~\eqref{eq:gub_induct} by supposing it holds for $i-1$. 
There are two cases to consider. 

\vspace{0.5em}
\noindent
\textit{Case 1. $\delta_{i-1} = 0$.} 
Since \eqref{eq:gub_induct} is true for $i-1$, we have
\begin{equation*}
    \sum_{j=1}^{i-1}{U_{i-1,j}} \geq \frac{p}{p+1}\sum_{j=1}^{i-1}{|\rho_{j+1}|} u_b.
\end{equation*}
Since $\sum_{j=1}^{i-1}{U_{ij}} \geq \sum_{j=1}^{i-1}{U_{i-1,j}}$, we have
\begin{align*}
    \sum_{j=1}^{i}{U_{ij}} &= \sum_{j=1}^{i-1}{U_{ij}} + U_{ii} \geq \frac{p}{p+1}\sum_{j=1}^{i-1}{|\rho_{j+1}|} u_b + U_{ii}\\
    &= \frac{p}{p+1} \left(\sum_{j=1}^{i}{|\rho_{j+1}|} - |\rho_{i+1}|\right) u_b + U_{ii}.
\end{align*}
Take \eqref{eq:U_ii_2} into the above inequality, \eqref{eq:gub_induct} follows.

\vspace{0.5em}
\noindent
\textit{Case 2. $\delta_{i-1} > 0$.} 
There are two subcases to consider.

\vspace{0.5em}
\noindent
\textit{Case 2.1. $\forall \tau_k \in \tau(\rho_i): u_{k} \geq u_b/(p+1)$.} By \eqref{eq:max_util}, there must be at least $p$ tasks assigned to $\rho_i$. Additionally, by \eqref{eq:u_0_2}, the sequential utilization of the first task assigned to $\rho_i$ is at least $u_b/(p+1) + \delta_{i-1}$. Therefore, we have 
\begin{equation*}
    u_{ii} \geq \frac{p}{p+1} u_b + \delta_{i-1}.
\end{equation*}
Take the above inequality and \eqref{eq:u_0_2} into \eqref{eq:U_ii}, we have
\begin{align*}
    U_{ii} &\geq \left(\frac{p}{p+1} u_b {+} \delta_{i-1}\right) |\rho_{i+1}| {+} \left(\frac{u_b}{p+1} {+} \delta_{i-1}\right) (|\rho_{i}| {-} |\rho_{i+1}|)\\
    &\geq \frac{p}{p+1} |\rho_{i+1}| u_b + \delta_{i-1} |\rho_{i}|.
\end{align*}
Combine this with~\eqref{eq:gub_induct} for $i-1$, we have
\begin{align*}
    &\sum_{j=1}^i{U_{ij}} \geq \sum_{j=1}^{i-1}{U_{i-1,j}} + U_{ii} \geq \frac{p}{p+1} |\rho_{i+1}| u_b + \delta_{i-1} |\rho_{i}|\\
    &+ \frac{p}{p+1}\sum_{j=1}^{i-1}{|\rho_{j+1}|} u_b - \delta_{i-1}|\rho_{i}| u_b = \frac{p}{p+1}\sum_{j=1}^i{|\rho_{j+1}|} u_b.
\end{align*}
Therefore, \eqref{eq:gub_induct} holds for Case 2.1. 

\vspace{0.5em}
\noindent
\textit{Case 2.2. $\exists \tau_k \in \tau(\rho_i): u_{k} < u_b/(p+1)$.}
This can happen only if $u_{i,i-1} + u_k > u_b$. By combining the above two inequalities, we have
$u_{i,i-1} + u_b/(p+1) > u_b$,
and thus
$$u_{i,i-1} > \frac{p}{p+1} u_b.$$
Combine it with $\delta_{i-1} = p/(p+1) u_b - u_{i-1,i-1}$, we have
$$u_{i,i-1} - u_{i-1,i-1} \geq \frac{p}{p+1} u_b - u_{i-1,i-1} = \delta_{i-1}.$$
Since $U_{i,i-1} - U_{i-1,i-1} \geq (u_{i,i-1} - u_{i-1,i-1}) |\rho_{i+1}|$, it follows
\begin{equation}
\label{eq:special_2}
    U_{i,i-1} - U_{i-1,i-1} \geq \delta_{i-1} |\rho_{i+1}|.
\end{equation}
Take \eqref{eq:u_ii} and \eqref{eq:u_0_2} into \eqref{eq:U_ii}, we have
\begin{align}  
\label{eq:U_ii_3}
    U_{ii} \geq \left(\frac{p}{p+1} u_b {-} \delta_i\right) |\rho_{i+1}| {+} \left(\frac{u_b}{p+1}  {+} \delta_{i-1}\right) (|\rho_{i}| {-} |\rho_{i+1}|).
\end{align}
Combine \eqref{eq:special_2} and \eqref{eq:U_ii_3} with~\eqref{eq:gub_induct} for $i-1$, we have
\begin{align*}
    &\sum_{j=1}^iU_{ij} = U_{ii} + \sum_{j=1}^{i-1}U_{i,j} = U_{ii} + U_{i,i-1} + \sum_{j=1}^{i-2}U_{i,j}\\
    &\geq U_{ii} + \sum_{j=1}^{i-1}U_{i-1,j} + (U_{i,i-1} - U_{i-1,i-1})\\
    &\geq  \left(\frac{p}{p+1} u_b - \delta_i\right) |\rho_{i+1}| + \left(\frac{u_b}{p+1} + \delta_{i-1}\right) (|\rho_{i}| - |\rho_{i+1}|)\\
    &~~~+ \frac{p}{p+1}\sum_{j=1}^{i-1}{|\rho_{j+1}|} u_b - \delta_{i-1}|\rho_{i}| + \delta_{i-1}|\rho_{i+1}|\\
    &\geq \frac{p}{p+1} \sum_{j=1}^i{|\rho_{j+1}|} u_b - \delta_i |\rho_{i+1}|.
\end{align*}
So~\eqref{eq:gub_induct} holds for Case 2.2. This concludes that \eqref{eq:gub_suf_cond} follows.

\section*{Acknowledgement}
Marco Caccamo was supported by an Alexander von Humboldt Professorship endowed by the German Federal Ministry of Education and Research.

\bibliographystyle{IEEEtran}
\bibliography{IEEEabrv,strict_partition}
\end{document}